\documentclass[useAMS,usenatbib]{mnras}
\usepackage{amsmath}
\usepackage{mathtools}
\usepackage{amssymb}
\usepackage{graphicx}
\usepackage{multirow}
\usepackage{color}
\usepackage{array}
\usepackage{graphicx}
\DeclareMathOperator*{\minimize}{minimize}
\DeclarePairedDelimiter\norm{\lVert}{\rVert}

\title[ML and light echoes around black holes]{Machine learning application to detect light echoes around black holes}

\author[P. Chainakun et al.]{P. Chainakun$^{1,2}$\thanks{E-mail: \href{mailto:pchainakun@g.sut.ac.th}{pchainakun@g.sut.ac.th}}, N. Mankatwit$^{1}$, P. Thongkonsing$^1$, A. J. Young$^3$ \\
$^1$School of Physics, Institute of Science, Suranaree University of Technology, Nakhon Ratchasima 30000, Thailand\\
$^2$Centre of Excellence in High Energy Physics and Astrophysics, Suranaree University of Technology, Nakhon Ratchasim 30000, Thailand\\
$^3$H. H. Wills Physics Laboratory, Tyndall Avenue, Bristol BS8 1TL, UK}
\date{Accepted XXX. Received YYY; in original form ZZZ}

\pubyear{2021}

\begin{document}
\label{firstpage}
\pagerange{\pageref{firstpage}--\pageref{lastpage}}
\maketitle

\begin{abstract}
X-ray reverberation has become a powerful tool to probe the disc-corona geometry near black holes. Here, we develop Machine Learning (ML) models to extract the X-ray reverberation features imprinted in the Power Spectral Density (PSD) of AGN. The machine is trained using simulated PSDs in the form of a simple power-law encoded with the relativistic echo features. Dictionary Learning and sparse coding algorithms are used for the PSD reconstruction, by transforming the noisy PSD to a representative sparse version. Then, the Support Vector Machine is employed to extract the interpretable reverberation features from the reconstructed PSD that holds the information of the source height. The results show that the accuracy of predicting the source height, $h$, is genuinely high and the misclassification is only found when $h>15 r_{\rm g}$. When the test PSD has a bending power-law shape, which is completely new to the machine, the accuracy is still high. Therefore, the ML model does not require the intrinsic shape of the PSD to be determined in advance. By focusing on the PSD parameter space observed in real AGN data, classification for $h \leq 10r_{\rm g}$ can be determined with 100\% accuracy, even using a PSD in an energy band that contains a reflection flux as low as $10\%$ of the total flux. For $h > 10r_{\rm g}$, the data, if misclassified, will have small uncertainties of $\Delta h \sim 2$--$4r_{\rm g}$. This work shows, as a proof of concept, that the ML technique could shape new methodological directions in the X-ray reverberation analysis.

\end{abstract}


\section{Introduction}

X-ray reverberation time delays arise due to the differences in the light travel time between the photons from the direct coronal emission and those which are observed after reflection from the accretion disc \citep[see][for a review]{Uttley2014}. The X-ray reverberation signatures in the lag-spectra have been previously investigated using either the standard lamp-post geometry where the X-ray source is assumed to be isotropic and point-like \citep[e.g.,][]{Wilkins2013, Cackett2014, Emmanoulopoulos2014, Chainakun2015, Chainakun2016, Epitropakis2016, Caballero2018, Ingram2019, Caballero2020} or the extended corona model \citep[e.g.,][]{Wilkins2016, Chainakun2017, Chainakun2019b}. The results suggested that the corona should be confined within $\sim 10r_{\rm g}$ from the central black hole, where $r_{\rm g}=GM_{\rm BH}/c^2$ is the gravitational radius, $G$ is the gravitational constant, $M_{\rm BH}$ is the black hole mass, and $c$ is the speed of light. The lag profiles, however, need to be produced and compared between two energy bands, usually one which is continuum dominated and the other which is reflection dominated. 

On the other hand, the use of the Power Spectral Density (PSD) showing the power of variability as a function of the temporal frequency allows us to probe the reverberation signatures using only a single energy band. An attempt at modelling reverberation signatures in the PSD profiles of AGN has been carried out by \cite{Epitropakis2016} and \cite{Chainakun2019a}. Due to the limitation of the signal to noise ratio, the reverberation features in the PSD may not be easily detected using current X-ray telescopes such as \emph{XMM-Newton}. To place strong constraints on the X-ray source height, the underlying intrinsic shape of the PSD onto which the reverberation is imprinted should be determined first \citep{Emmanoulopoulos2016}. Furthermore, if the corona is extended, the reverberation features may be even more subtle \citep{Chainakun2019a}. It is, therefore, interesting to see if these reverberation echo signatures that appear in the PSD profiles can be directly discriminated and recognized through the proper use of machine learning algorithms.

Dictionary-based machine learning methods are widely used in extracting and reconstructing signals \citep[e.g.,][]{Pati1993, Olshausen1996}. The dictionary could be simply created from formal mathematical functions and in this case it is generally called a fixed dictionary. Alternatively, the dictionary could be learned from the training data through the use of a Dictionary Learning (DL) algorithm. The DL technique can be applied to detect characteristic features in a given data set by deriving a finite collection of dictionary elements (referred to as atoms). These atoms represent local, small recurring patterns imprinted within the data so that a whole dictionary becomes a compact representation of a large scale data set. One of the algorithms that could learn the dictionary is known as sparse coding. It relies on the assumption that any new signals can be reconstructed using a few sets of atoms in the dictionary. The approximated, reconstructed signal then can be obtained by encoding the input through a linear combination of the atoms and can produce a smooth, or sparse, version of the original signal. 

Here, we are interested in using DL and sparse coding to reconstruct the PSD (e.g, to produce the sparse version of a noisy PSD). DL as a machine learning tool has been used in the past to analyze astronomical data. For example, \cite{Lachowicz2010} studied the evolution of quasi-periodic oscillation (QPO) in black hole binaries and found that the DL technique could provide clearer description of the low-frequency QPO which is composed of multiple independent oscillations with constant frequency, compared to the standard technique. \cite{Pieringer2019} developed a DL-based algorithm to visualize relevant patterns in light curves of variable stars. To the best of our knowledge, the application of DL for PSD reconstruction and reverberation detection in AGN has not yet been studied and reported. 

After PSD reconstruction phase, we apply the Support Vector Machine (SVM) \citep[e.g.,][]{Boser1992, Cortes1995}, one of the most robust and commonly used prediction methods, to analyze the sparse PSD data and predict X-ray source locations. We note that the term ``prediction" here actually means we are predicting the discrete values, since the SVM is one of the classification based models. Very high quality data and models would be required to train a model to predict continuous values (regression problems). Given that degeneracies are often found in the timing models and the intrinsic shape of the PSD is not usually known in advance \citep{Papadakis2016, Emmanoulopoulos2016, Chainakun2019a}, sufficiently detailed data may not be available. Therefore, we first choose to use the SVM model so that the X-ray source locations can be classified into different ranges of source heights, which can be straightforwardly fine-tuned to satisfy the accuracy of the model predictions.

The problem statement here is whether or not the incorporated DL and SVM models, trained using simple power-law PSD data, can identify the reverberation features on more complex PSD shapes (e.g., bending power-law), and predict accurate source heights. This is a test of how well the machine can adapt to explain the new data. Also, it is for evaluating whether or not the accurate prediction of the source height requires the intrinsic shape of the PSD to be determined in advance. 

The paper is structured into seven sections and is framed by the opportunity and challenge to use ML as a new probe for the X-ray source location. The reverberation model development is explained in Section 2, including how we compute the PSD and encode the reverberation echo signatures. Data preparation for training and testing the machine is described in Section 3. In Section 4, we present in details the DL and SVM algorithms associated with our ML model. Evaluations of how the ML model performs are presented in Section 5. The results are discussed in Section 6, following by the conclusion in Section 7.

\section{Reverberation model}

\subsection{PSD and echo response}
We assume a lamp-post geometry where an accretion disc is illuminated by a point X-ray source located at a height $h$ on the symmetry axis above the black hole. The variability power from the X-ray source is modelled in term of the power spectral density (PSD) written in the form of a simple power-law,
\begin{equation}
P_{\rm 0}(f,E_{j}) \propto f^{-\gamma} \; ,
    \label{psd0}
\end{equation} 
where $f$ is the frequency, $\gamma$ is the index of the power-law and $E_{j}$ is the energy band of interest. 

We use the {\sc kynxilrev}\footnote{\url{https://projects.asu.cas.cz/stronggravity/kynreverb}} model \citep{Caballero2020} to compute the ionised reflection and reverberation responses from the accretion disc under the lamp-post configuration, with the source height ($h$) and the inclination angle ($i$) being model parameters. Note that the {\sc kynxilrev} is a part of {\sc kyn}\footnote{\url{https://projects.asu.cas.cz/stronggravity/kyn}} models introduced by \cite{Dovciak2004a} and \cite{Dovciak2004b}. The {\sc kynxilrev} model performs the ray-tracing of photons along the Kerr geodesics and computes the reflection, or re-processing X-rays from the ionised disc using the {\sc xillver} table model \citep{Garcia2010, Garcia2013} taking into account all relativistic effects due to the light propagation.

In {\sc kynxilrev}, we also fixed the black hole mass and the spin parameter to be $M=10^{6}M_{\odot}$ and $a=1$, respectively. The disc is set to be a standard geometrically thin optically thick disc \citep{Shakura1973} that extends from inner-most stable circular orbit to the outer radius of $r_{\rm out}=10^3r_{\rm g}$. The X-ray continuum is in the form of a cut-off power-law: $F(E) \propto E^{-\Gamma}{\rm e}^{-E/E_{\rm c}}$ where the energy cut-off is fixed at $E_{\rm c}=300$~keV. Other parameters in {\sc kynxilrev}, if not mentioned, are set at their default values.

\begin{figure}
    \centerline{
        \includegraphics[width=0.45\textwidth]{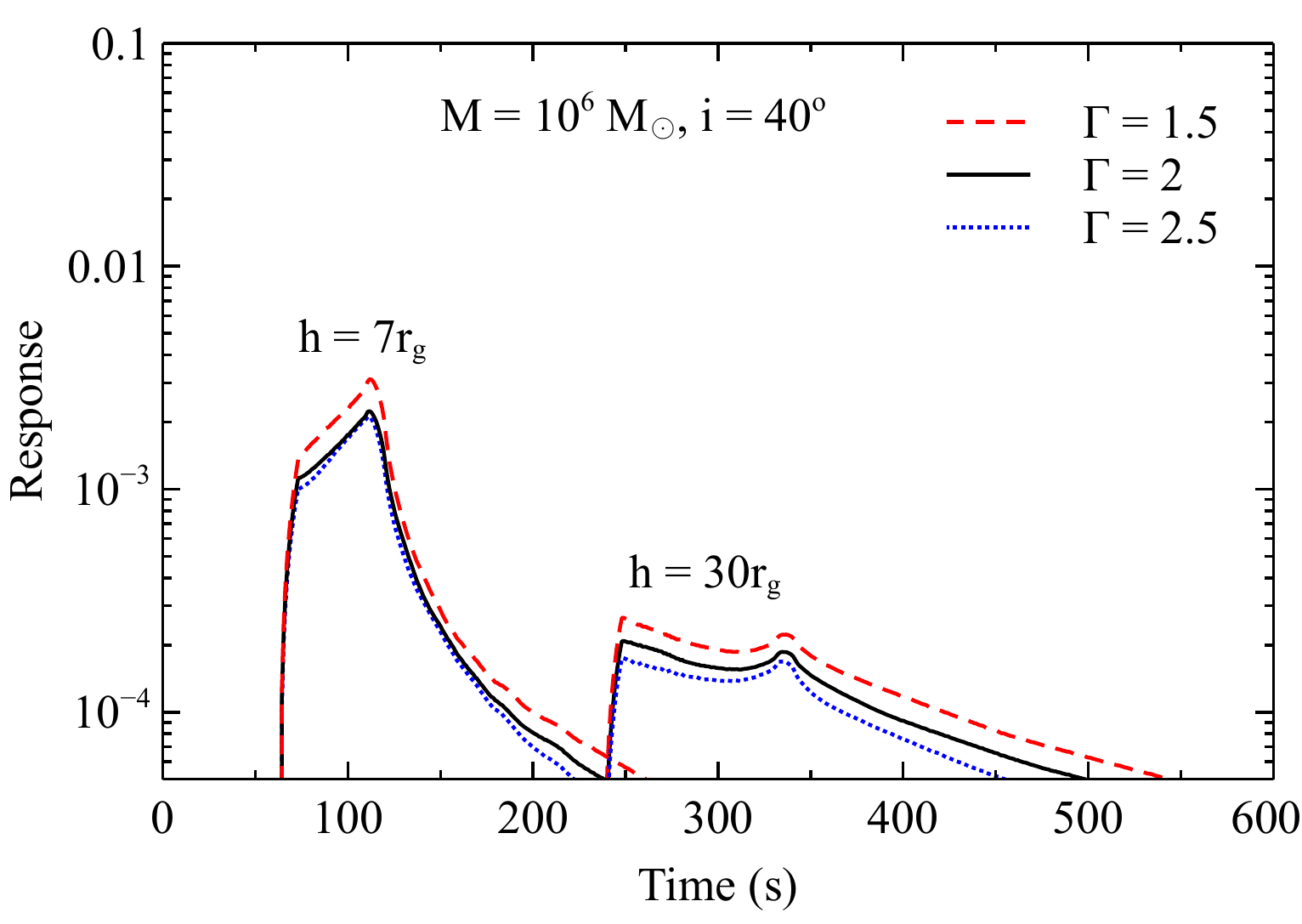}
    }
    \vspace{0.2cm}
    \centerline{
        \includegraphics[width=0.45\textwidth]{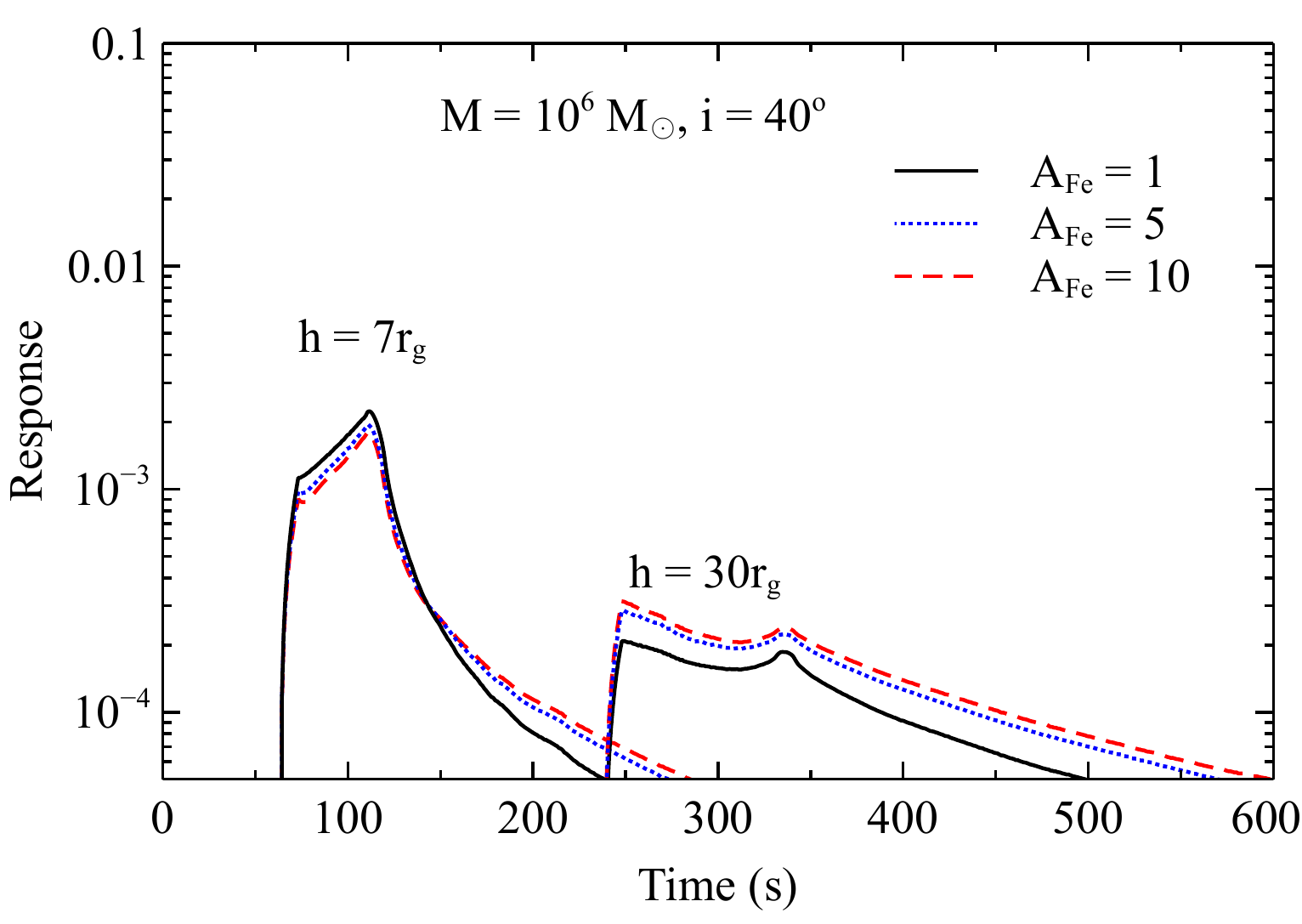}
    }
    \caption{Relativistic response functions in 5--7 keV band when $h=7r_{\rm g}$ and $30r_{\rm g}$ produced by {\sc kynxilrev} model varying with the power-law energy index of the primary flux, $\Gamma$, (top panel) and the iron abundance of the disc material, $A_{\rm Fe}$ (bottom panel). Here we assume a radial power-law density profile for the disc. Other parameters if not stated are frozen at $a=1$, $M=10^{6}M_{\odot}$, $A_{\rm Fe}=1$, and $\gamma = 2$.}
    \label{fig1}
\end{figure}

We execute the {\sc kynxilrev} model outside {\sc XSPEC} and produce the light curves for the observed reflection in the energy bands of interest. These reflection light curves describe the observed amount of reflected flux from the accretion disc as a function of time due to an instantaneous flash of the X-ray source, and hence represent the full-relativistic disc response functions, $\psi(t, E_{j})$. Dependence of reverberation responses in 5--7~keV band on the the source height, $h$, photon index of the X-ray continuum, $\Gamma$, and the iron abundance, $A_{\rm Fe}$, produced by the {\sc kynxilrev} model are presented as an example in Fig.~\ref{fig1}. Among these parameters, the source height has relatively and significantly high impact on the response functions, in agreement with previous literature \citep[e.g.,][]{Cackett2014, Emmanoulopoulos2014, Chainakun2015, Epitropakis2016}. 

\subsection{Reverberation features on the PSD}
We consider two X-ray components including direct continuum and reflection X-rays which can be observed by a distant observer. The lightcurves can be formalized as a linear additive mixture of direct continuum flux and the reverberation flux. The total response of the system then can be written as the sum of contributions from both components. The modulus squared of the Fourier transform of the total response, $\Psi(f, E_{j})$, could produce the observed PSD, $P_{\rm obs}(f,E_{j})$, by using $P_{\rm 0}(f,E_{j})$ from eq.~\ref{psd0} as an input PSD \citep[e.g.,][]{Uttley2014, Papadakis2016, Chainakun2019a}: 
\begin{equation}
P_{\rm obs}(f,E_{j}) \propto |\Psi(f,E_{j})|^{2} P_{\rm 0}(f,E_{j}) \; .
    \label{psd}
\end{equation} 
The response function then acts as a filter on the input $P_{\rm 0}(f,E_{j})$, producing the observed PSD with a prominent dip following by oscillatory behaviour with smaller amplitude at higher frequencies \citep{Papadakis2016, Chainakun2019a}. These reverberation echo features are independent of the driving signals but are more prominent in a band that is more dominated by reflection. The impulse responses function varies significantly with the the geometry of the system, and so do the reverberation echo features imprinted on the $P_{\rm obs}(f,E_{j})$.

Note that the key information of the source geometry is imprinted in reverberation features produced by the response function, but the clean reverberation signal is never observable. The observer always sees a mixture of flux contributed by the direct continuum emission and X-ray reverberation. The goal of this work is to develop an ML model that helps separate the reverberation features in order to infer the corresponding source location. 

\section{Data preparation and simplification}

Firstly, the data need to be properly prepared and normalised so they are as simple as possible while retaining essential important information (e.g., reverberation features) for the machine to learn. Since the effect of the energy band is likely to only dilute the reverberation features, we do not model the PSD of all energy bands. Instead, we produce the response function only in one particular band (Section 2.1--2.2) and employ the reflection fraction $R_{\rm f}$ defined as (reflection flux)/(continuum + reflection) to factorize energy-band dependent shapes of the PSD. Lower and higher values of $R_{\rm f}$ refer to the energy bands that are more dominated by continuum and by reflection, respectively. The $R_{\rm f}$ varies between 0 and 1 where the lower bound 0 and upper bound 1 means only continuum and only reflection flux contributes to the light curve, respectively. $R_{\rm f}=0.5$ means continuum and reflection flux are equally contributed. There is no preferred value of $R_{\rm f}$ before we start to test the machine, so it is varied to be 0.1, 0.3, 0.5, 0.7, and 0.9. The model grid of $R_{\rm f}$ can be adjusted later based on the performance of the machine.

We simulate the PSD data using eqs.~\ref{psd0}--\ref{psd} by varying the power-law index $\gamma$ between 1--2 with equal spacing of 0.1. We produce 75 response functions in total using the {\sc kynxilrev} model \citep{Caballero2020} by selecting an ensemble of 15 source heights $h \in \{ 2.3, 4, 6, 8, 10, 12, 14, 16, 18, 20, 22, 24, 26, 28, 30 \} r_{\rm g}$ and 5 inclinations $i \in \{ 5^{\circ}, 30^{\circ} ,45^{\circ}, 60^{\circ}, 75^{\circ} \}$. The modulus squared of the Fourier transform of each response function is used to modify each original PSD signal and hence 3,750 reverberation-encoded PSD signals are obtained, each of which is produced in 512 frequency bins. These PSD data are termed ``clean data" meaning that the their profiles are smooth and are simulated using the simple power-law model. The prepared clean data then are fed through the workflow that consists of the DL and SVM algorithms, which will be described in detail in the next Section. 

\section{Machine learning algorithm}
\begin{figure}
    \centerline{
        \includegraphics[width=0.5\textwidth]{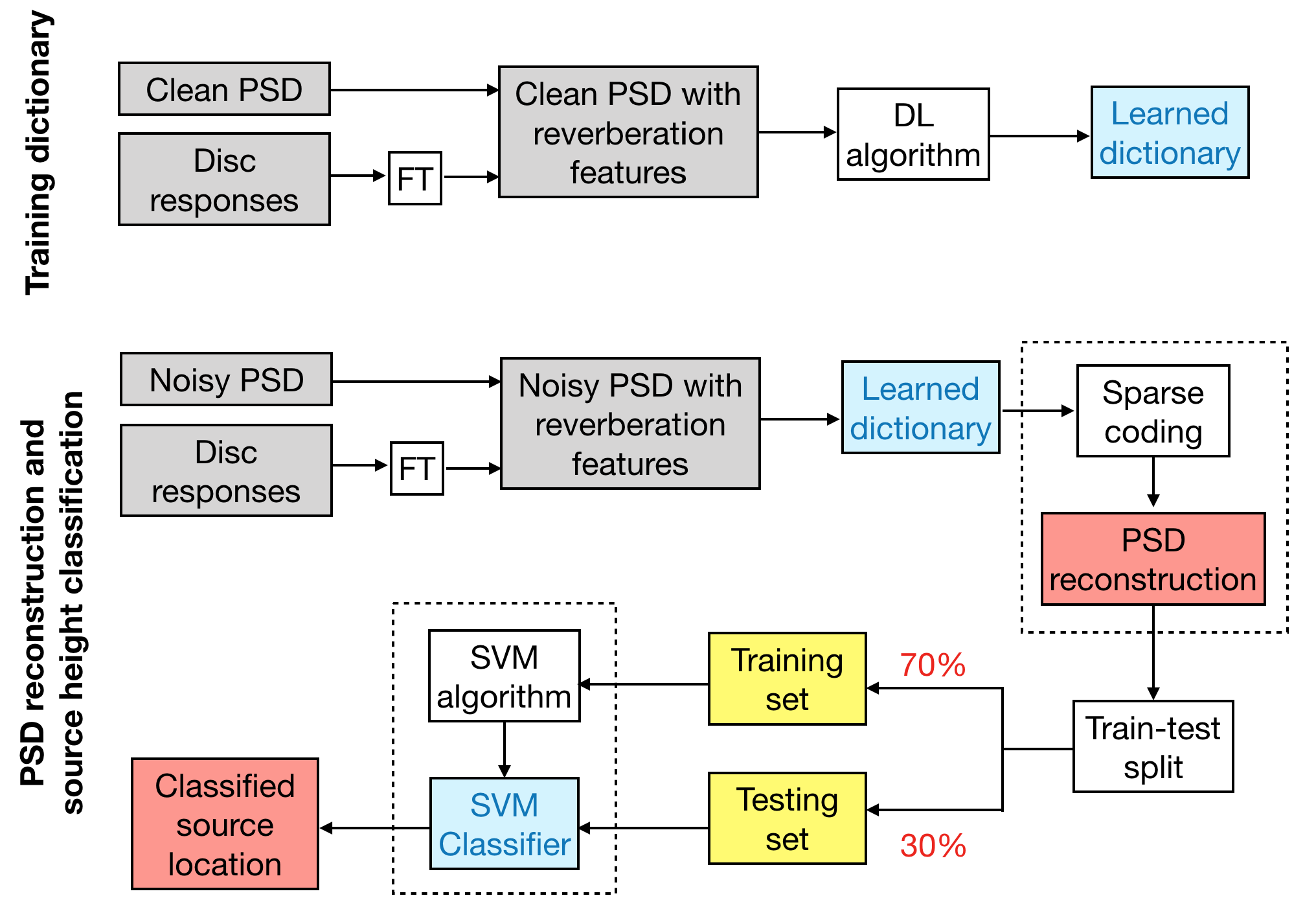}
    }
    \caption{Flowchart representing our ML algorithm. The PSD data, in the form of a power-law with reverberation features added using the method outlined in Section 2, are used to train the DL to obtain the learned dictionary. Noisy PSD data are produced via the method of \protect\cite{Timmer1995} and are split into training set (70\%) and testing set (30\%). The DL and sparse coding algorithms are used to reconstruct the PSD first, before their characteristic echo features are extracted and classified using the SVM model. Each predicted class relates to a specific range of source heights. Cross validation is preformed during the training phase while the test data are held out for final evaluation. See text for more details.}
    \label{fig-flowchart}
\end{figure}

The machine is trained step by step using Dictionary-based Learning (DL) and Support Vector Machine (SVM) algorithms. Our goal is to use machine trained by DL and SVM to reconstruct the PSD signal with reverberation echo features and then predict the location of the X-ray source. Note that all ML models and supported functions used here are from {\sc scikit-learn}\footnote{\url{https://scikit-learn.org/}} \citep{Pedregosa2012}. An illustrative chart of these processes is presented in Fig.~\ref{fig-flowchart}.

\subsection{Dictionary Learning and sparse coding}

Once 3,750 PSD data are prepared as described in Section 3, we feed them into the DL algorithm. Mathematically speaking, these data are denoted by a column-wise matrix
\begin{equation}
\textbf{X} = [\textbf{x}_{1}, \textbf{x}_{2}, \textbf{x}_{3},...,\textbf{x}_{N}] \in \mathbb{R}^{M \times N} \;,
    \label{eq:X}
\end{equation} 
where $N=3,750$ is the total number of PSDs and $M=512$ is the size of each PSD (i.e., the number of temporal frequency bins). A dictionary is prepared by specifying a number of atoms, $K$, representing a number of elements inside the dictionary. Each atom is a vector of size $M$ that records local, recurring patterns imprinted within the training data so that a whole dictionary with $K$ atoms can become a compact representation of a large scale data set. 

The representative dictionary consisting of $K$ columns of atoms can be written as
\begin{equation}
\textbf{D} = [\textbf{d}_{1}, \textbf{d}_{2}, \textbf{d}_{3},...,\textbf{d}_{K}] \in \mathbb{R}^{M \times K} \;,
    \label{eq:D}
\end{equation} 
where $ \textbf{d}_k \in \mathbb{R}^{M}$ and $K<N$. An algorithm we employ for training the DL is the sparse coding, whose matrix is given by
\begin{equation}
\textbf{C} = [\textbf{c}_{1}, \textbf{c}_{2}, \textbf{c}_{3},...,\textbf{c}_{N}] \in \mathbb{R}^{K \times N} \;.
    \label{eq:C}
\end{equation} 

To find the best dictionary is to factorize the data matrix \textbf{X} into the representative dictionary matrix \textbf{D} and coding matrix \textbf{C}, leading to the joint optimization  problem between \textbf{D} and \textbf{C}:
\begin{equation}
     \minimize_{\textbf{D,C}} \, \norm{\textbf{X} - \textbf{DC}}^{2}_{F} + \lambda\norm{\textbf{C}}_{l_{r}} \;,
    \label{eq:tp}
\end{equation}
which is executed under the constraints
\begin{equation}
    \norm{\; \textbf{d}_{k} \;}_{F}^{2} = 1 \; {\rm for \; all} \; 0 \leq k < K \;.
    \label{eq:constraints}
\end{equation} 
Note that \textbf{d}$_{k}$ is the $k^{\rm th}$ atom inside the dictionary \textbf{D} and $K$ is the total number of atoms we want to learn. $\norm{\; . \;}_{F}^{2}$ denotes the squared Frobenius matrix norm which is the sum of the squares of all the matrix entries. The first term in eq.~\ref{eq:tp} ensures \textbf{D} is good at representing the data \textbf{X} while the second term ensures that only few entries in \textbf{D} will be activated in recovery of \textbf{D}. This problem is solved by fixing one of the two variables (\textbf{D} or \textbf{C}) while optimizing the other one. The regulization parameter $\lambda$ is a sparsity controlling parameter and is set to 1, as default. The parameter $l_{r}$ is the penalty factor for the sparse solution $\textbf{c}_k$. The $l_{0}$ refers to the solution obtained via the Orthogonal Matching Pursuit (OMP) based on a greedy approach \citep{Pati1993}. For $l_{1}$, the minimization fits the sparse coding solution that uses Lasso regression for variables selection and regularization \citep{Olshausen1996}. 

We use {\tt sklearn.decomposition.DictionaryLearning()} for training the machine via DL method with both OMP and Lasso algorithms \citep{Pedregosa2012}. In this way, new data $\textbf{X}_{\rm new}$ can be reconstructed using the learned dictionary via a linear combination
\begin{equation}
{\textbf X}_{\rm new} \sim {\textbf D} {\textbf C}_{\rm new}\;,
    \label{eq:Xnew}
\end{equation} 
where $\textbf{C}_{\rm new}$ can be thought as sparsity-inducing coefficients (codes) that activated specific trained features inside the specific atoms of the dictionary. The Reconstruction phase is performed using {\tt sklearn.decomposition.SparseCoder()}. Note that we train the machine through the DL algorithm using only the clean PSD data in the form of a power law. This is because we would like the machine to reconstruct the PSD in the way as to produce sparse version of the observed PSD which usually shows scatter.

Dictionaries with a fixed number of atoms $K$ = 12, 16, 24, 32, 64, 96, 128, and 256 are investigated. The reconstruction PSD profiles produced using OMP and Lasso algorithms are compared. The other parameters associating with the ML functions, if not stated, are set at their default values.

\subsection{Support Vector Machine}

\begin{figure}
    \centerline{
        \includegraphics[width=0.45\textwidth]{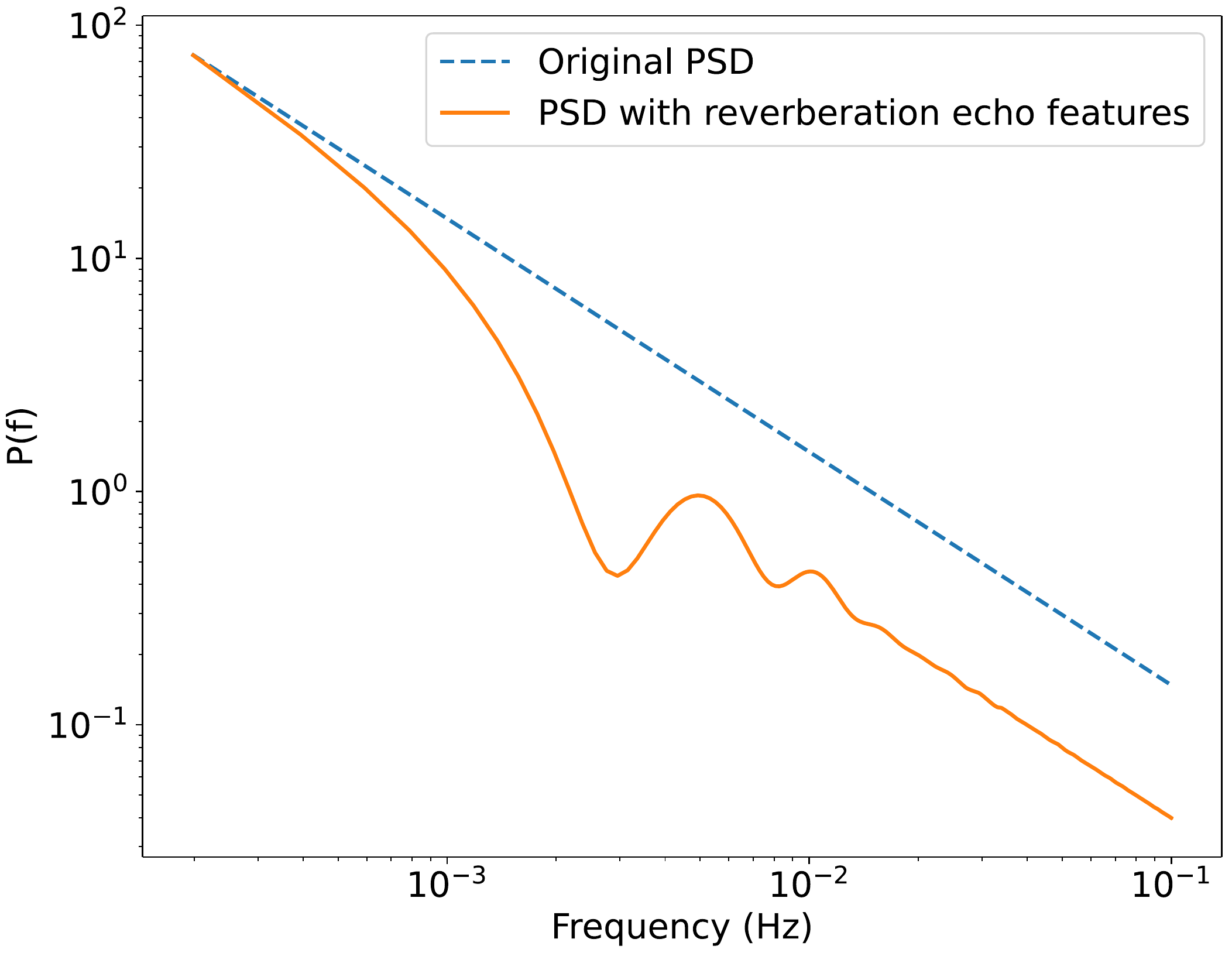}
    }
    \vspace{0.2cm}
    \centerline{
        \includegraphics[width=0.45\textwidth]{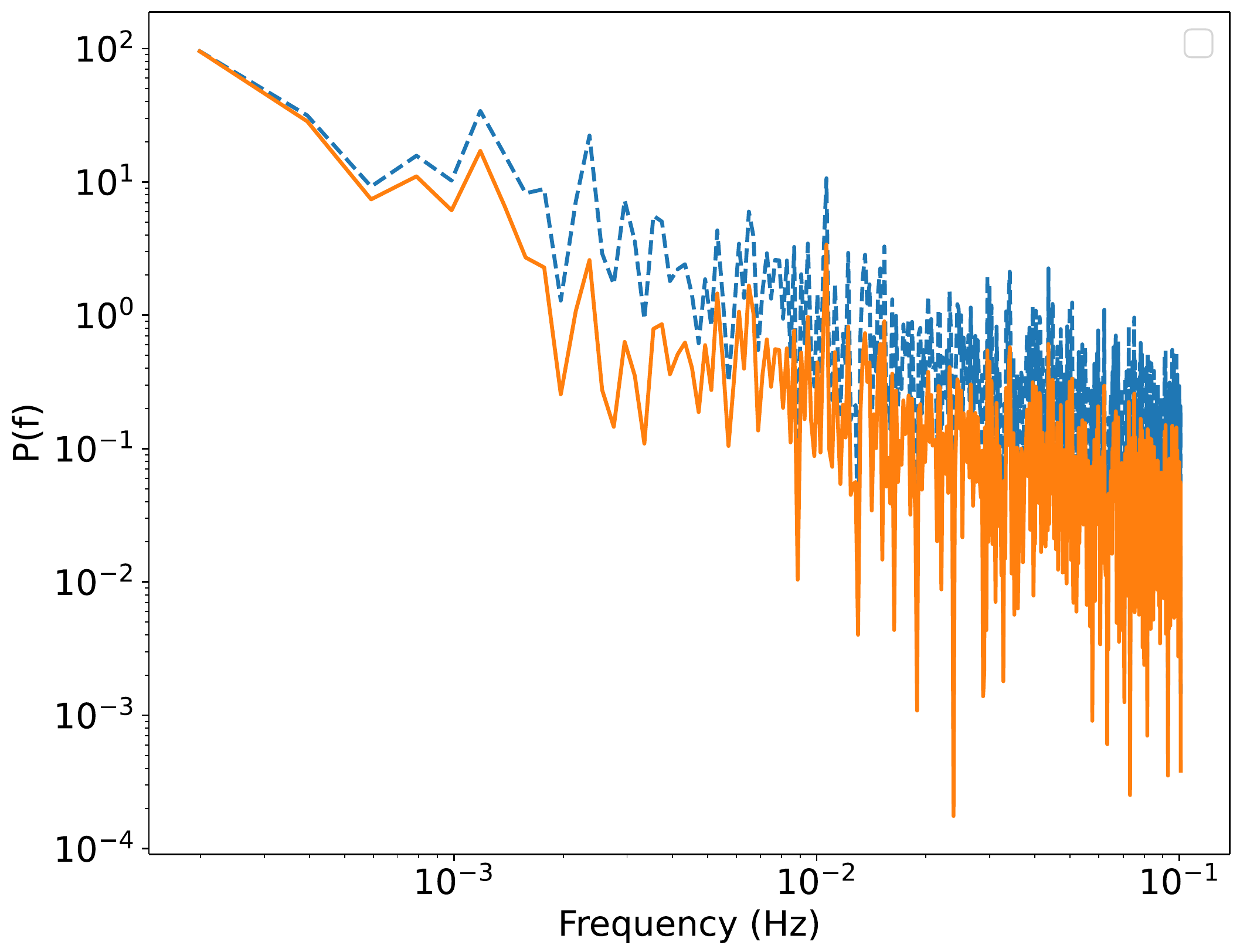}
    }
    \caption{Top panel: Clean PSD in the form of a power-law. Bottom panel: Noisy PSD generated using the method of \protect\cite{Timmer1995}. Blue dashed-lines show the original simple PSD while orange solid-lines show the associating PSD encoded with reverberation echoes produced by the X-ray source at $h=14r_{\rm g}$.}
    \label{PSD_training}
\end{figure}

We simulate the noisy PSD profiles following \cite{Timmer1995} and convolve them with disc response functions for 15 different source heights and 5 inclination angles. Examples are shown in Fig.~\ref{PSD_training}. The noisy PSD with reverberation echo features are reconstructed using the learned dictionary. Then, they are split randomly using {\tt sklearn.model\_selection.train\_test\_split()} that is available in {\sc scikit-learn} \citep{Pedregosa2012} with the training/testing sample ratio of 70/30 (i.e., 70\% is used for training the machine while the remaining 30\% is kept for testing and evaluating the performance of ML). These PSD data, after being reconstructed and split, are used for training and testing the Support Vector Machine (SVM) technique \citep{Boser1992, Cortes1995}.

The aim of the SVM algorithm is to find a hyperplane in an N-dimensional space (N = number of features) that distinctly classifies the data points. Data falling on different sides of the hyperplane are attributed to different classes. The best hyperplane is the one that maximizes the separation distance between the closest data points of different classes. It can be obtained by maximizing the width of the margin given by $2/\norm{w}$ where $w$ is a normal vector perpendicular to the hyperplane. To maximize the margin ($2/\norm{w}$), we implement the algorithms that minimize $\norm{w}$: 
\begin{equation*}
\begin{aligned}
& \underset{w,b,\zeta}{\text{minimize}}
& & \frac{1}{2} \norm{w}^2 + C \sum_{i=1}^{n} \zeta_i  \\
& \text{subject to}
& &  y_{i}\big{(} w^{T}\phi(x_i) + b \big{)} \geq  1 - \zeta_{i}, \\  
& & & \zeta_i \geq 0, \; i = 1, 2, 3,\ldots, n,
\end{aligned}
\label{eq:svm}
\end{equation*}
where $\phi(x_{i})$ is output kernel mapping of training data $x_i$, $\norm{w}^2 = w^{T}w$, and $C$ is the regularization parameter to control how much we incur a penalty when a training sample is misclassified. The parameter $\zeta_i$ measures the distance of the data from their correct margin boundary.  

Most simple type SVM is for binary classification, dividing data into two classes, e.g., 1 and 0. Considering only the source height, our PSD data have an ensemble of 15 different source heights, and hence fall maximally into 15 classes. In this multi-class classification, the problem could be broken down to multiple binary classifications, which is called one-vs-one (ovo). Kernel functions are used to lay hyperplanes of diverse shapes through the multi-class data points by mapping the input data into a higher-dimensional feature space. The key kernel functions investigated here are linear and radial basis functions (RBF). 

For comparison, the PSD data are further binned into 8, 16, 32, 64, 128 and 256 temporal frequency bins, from the original size of 512 bins. We normalize each of them by taking a logarithm of the variability power in each temporal frequency bin. The SVM for Support Vector Classification (SVC) is performed using {\tt sklearn.svm.svc()} with ovo decision function. In this case, the variability power is the feature for the machine to extract. The number of features then are the number of frequency bins we used. The given labels are the assigned source-height classes. The SVM parameters not mentioned here are kept as default. The accuracy of the SVM is evaluated through the use of {\tt sklearn.svm.svc.score()} that returns the mean accuracy on our test data and labels. 

\subsection{Hyperparameter tuning and cross-validation}

In SVM, the best choice of kernel functions depends on the nature of the data. There are also some parameters, known as hyperparameters, that exhibit their importance to the ML improvement and cannot be directly learned within estimators. Using the linear kernel, we need to optimize the hyperparameter $C$ which is the penalty parameter (regularization) controlling the trade-off between decision boundary and misclassification term. Higher $C$ means the data points may be classified more correctly, but also higher chance to overfit. In the RBF kernel, there is an additional hyperparameter $\Gamma$ which is a tuning parameter accounting for the smoothness of the decision boundary. This parameter defines the distance of influence of a single training point. Higher $\Gamma$ means closer points to the hyperplane will be considered to get the decision boundary (i.e., the points need to be closer to be considered in the same class).

The best hyperparameters can be fine-tuned using {\tt sklearn.model\_selection.GridSearchCV()}, by producing a grid of hyperparameters and evaluating the cross-validation among the group of data in the training set. The Stratified $k$-Folds cross-validator is used to split the training data into $k$ smaller sets with the same ratio of samples for each class across each split. Then, the model is trained using $k-1$ of the folds while using the other fold for validation, holding out the test data to still be completely unseen for final evaluation. This process helps preventing the model to have a perfect classification on training data set that it has just seen but fails to make accurate predictions on yet-unseen test data.

To reduce the number of free parameters, we fixed $k=5$ so that the training data is splitted into 5 folds having 4 folds (80\%) for training and 1 fold (20\%) for validation, and the process is repeated 5 times with 5 different training and validation set. Every fold then appears in the training set 4 times and the validation set once. Finally, a new model is built according to the best parameters obtained after the cross-validation process and is used to make predictions of the unseen test set for final evaluation. 

\section{Results and analysis}

\subsection{PSD reconstruction}

Several scenarios of training and testing are performed and evaluated. We find that the ML model prefers 64 temporal-frequency bins, meaning that we have 64 features in the classification model. The learned dictionary with 24 atoms and OMP decoding algorithm can provide the reconstructed PSD with the best accuracy on the SVM classification. Smaller number of atoms results in less accuracy, but larger number of atoms than 24 becomes more time consuming while the ML accuracy is not significantly improved. Therefore, we select to use $K$ = 24 for further analysis.

\begin{figure}
    \centerline{
        \includegraphics[width=0.5\textwidth]{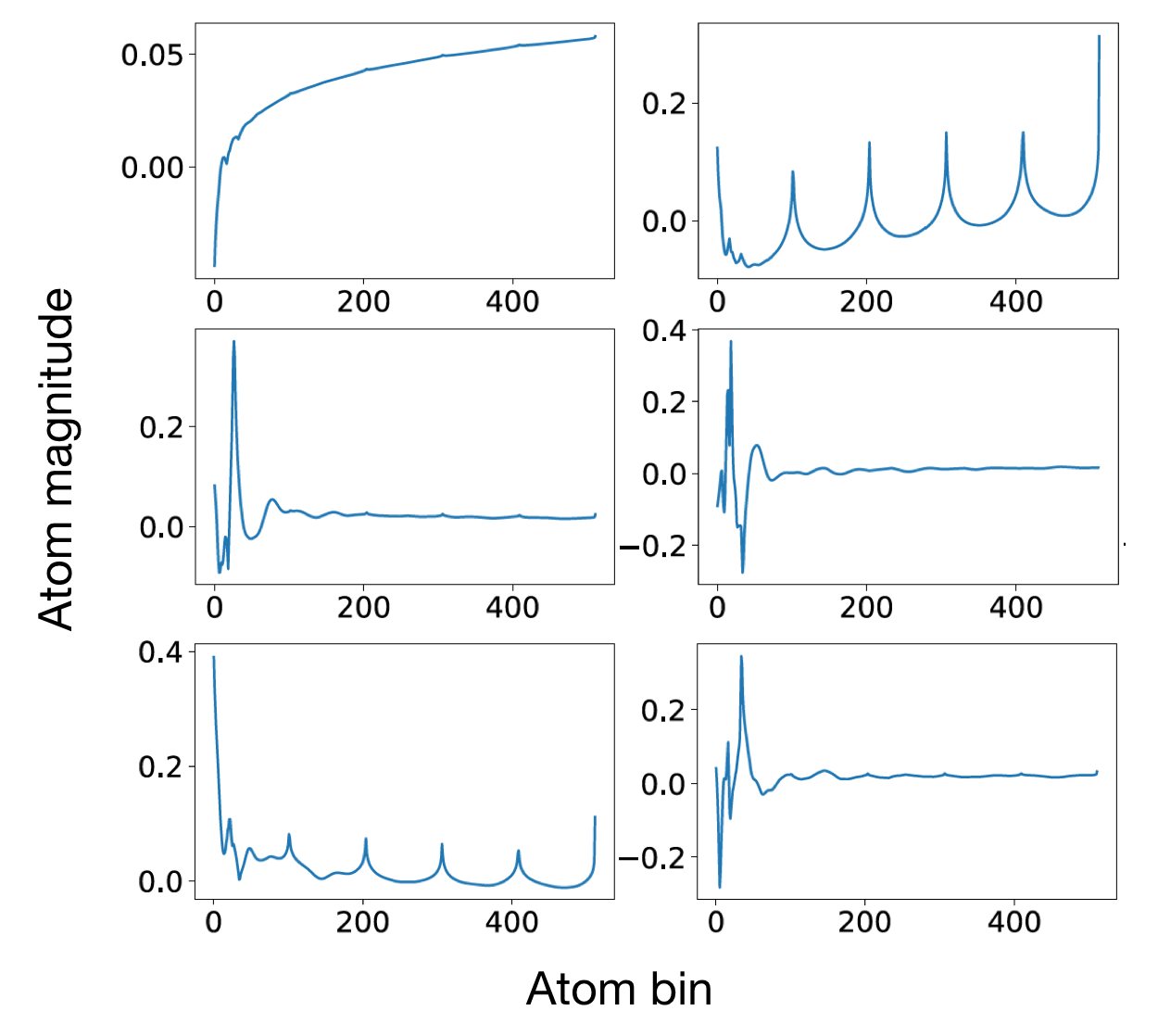}
    }
    \caption{Six randomly selected atoms from the learned dictionary. These atoms show local recurring patterns which are representative of those in large scale data set.
    }
    \label{fig-atoms}
\end{figure}

\begin{figure}
    \centerline{
        \includegraphics[width=0.5\textwidth]{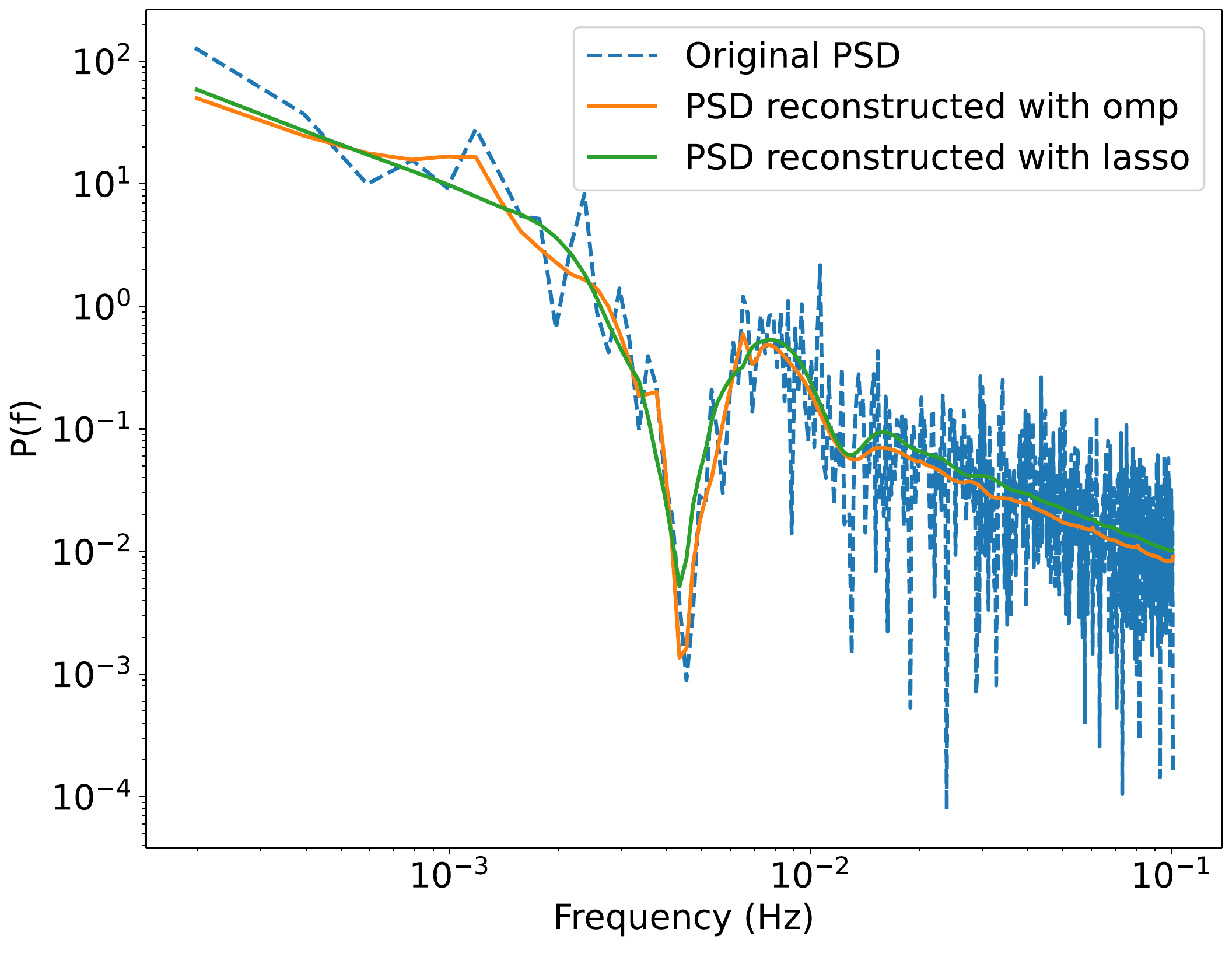}
    }
    \caption{PSD encoded with the reverberation features produced by the X-ray source at $h=6r_{\rm g}$. The noisy profile is shown in the blue dashed-line while the reconstruction ones obtained via DL and sparse coding using the OMP and the Lasso algorithms are shown in orange and green solid-lines, respectively.}
    \label{fig-PSD-recon}
\end{figure}

Fig.~\ref{fig-atoms} presents examples of the dictionary elements, or atoms, found during the dictionary learning. Visual pattern of the atoms should relate with the training PSD where reverberation features are encoded. Note that once the DL model is obtained, we produce a set of noisy signals via the method of \cite{Timmer1995}. The noisy PSD is reconstructed using the DL and sparse coding that activates a set of atoms to be produced as a representative sparse version of the original PSD. The example of reconstructed PSD using the OMP and the Lasso algorithm is shown in Fig.~\ref{fig-PSD-recon}. We can observe a fairly sparse reconstruction of the noisy signal, suggesting that the DL and sparse coding algorithms are capable of reconstructing the descriptive, clean features from the training data.

\subsection{Classifying the source height}

The PSD, after being reconstructed via DL and sparse coding, are split into 70\% and 30\% for traning and testing via the SVM algorithm. We investigate the performance of SVM with linear and RBF kernels on distinguishing the reconstruced reverberation patterns on the PSD data. Note that the kernel is a mapping function that transforms the data from one to a new space so that the SVM can better make the hyperplane decision boundary for its classification. There is one hyperparameter ($C$) for linear kernel and two hyperparameters ($C$ amd $\Gamma$) for RBF kernel. These parameters need to be fine-tuned and cannot be directly learned during the training process. Although their mathematical approaches are different, both kernels provide comparable high accuracy up to $\sim 1$. Due to its smaller number of hyperparameters and being less time consuming, we choose to investigate and present the results only in the case of SVM with a linear kernel.

\begin{figure}
    \centerline{
        \includegraphics[width=0.5\textwidth]{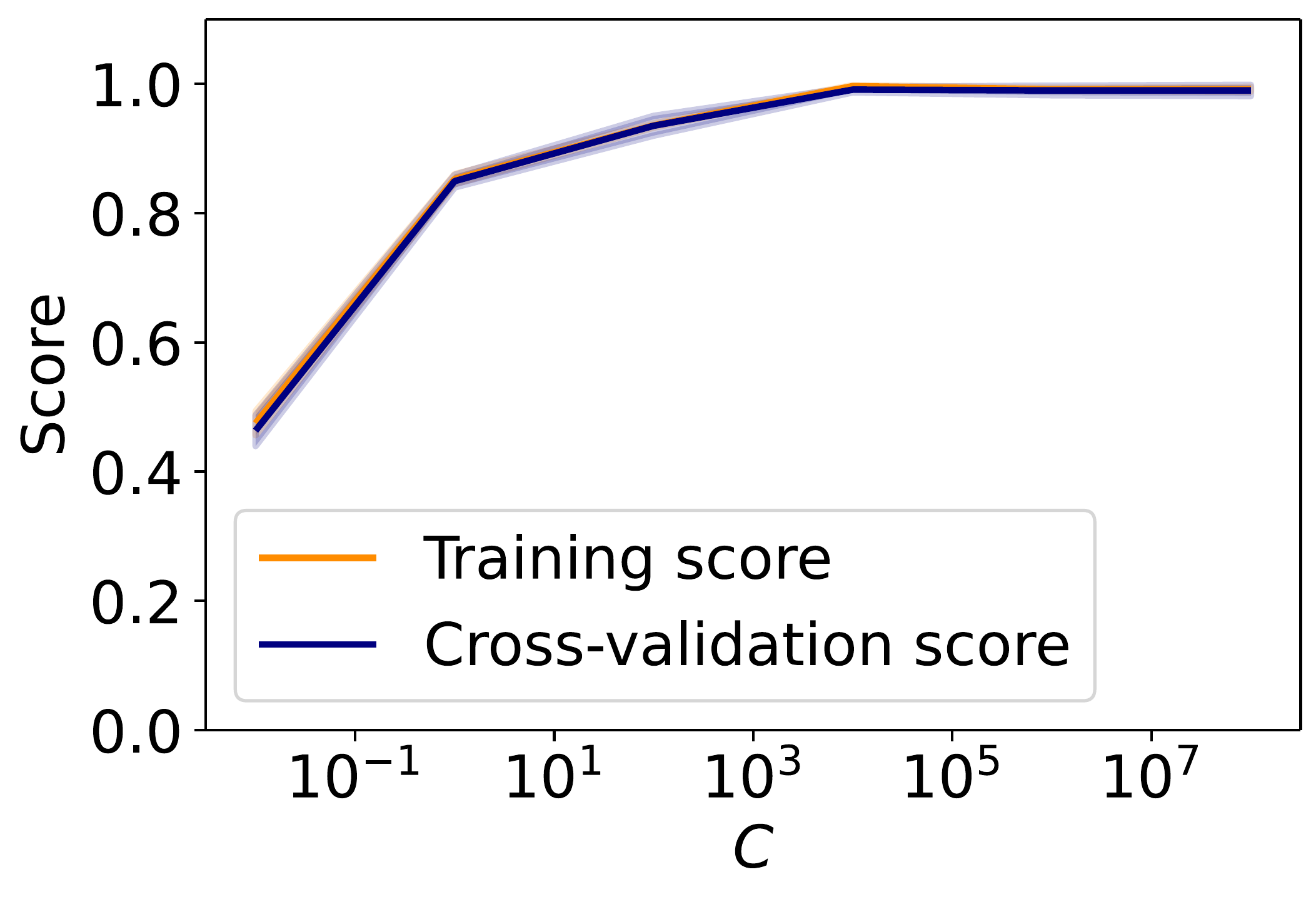}
    }
    \caption{Training scores (orange line) and validation scores (blue line) for classifying $h$ for different values of the hyperparameter $C$. The SVM classifier performs very well when $C \gtrsim 10^4$.}
    \label{fig-C-h}
\end{figure}

Fig.~\ref{fig-C-h} shows the results obtained from tuning the hyperparameter $C$ in the linear kernel using $k$-Folds cross-validator implemented in {\tt sklearn.model\_selection.GridSearchCV()}, as described in Section 4.3. Note that we set $k=5$ which means 20\% of training data is cross-validated to compute the validation error. It can be seen that the model is underfitting for the low values of $C$ since both training and validation scores are low. When $C\gtrsim 10^4$, the SVM classifier performs fairly well since both training score and validation scores are significantly high with the highest mean score of $\sim 1$.

\begin{figure}
    \centerline{
        \includegraphics[width=0.5\textwidth]{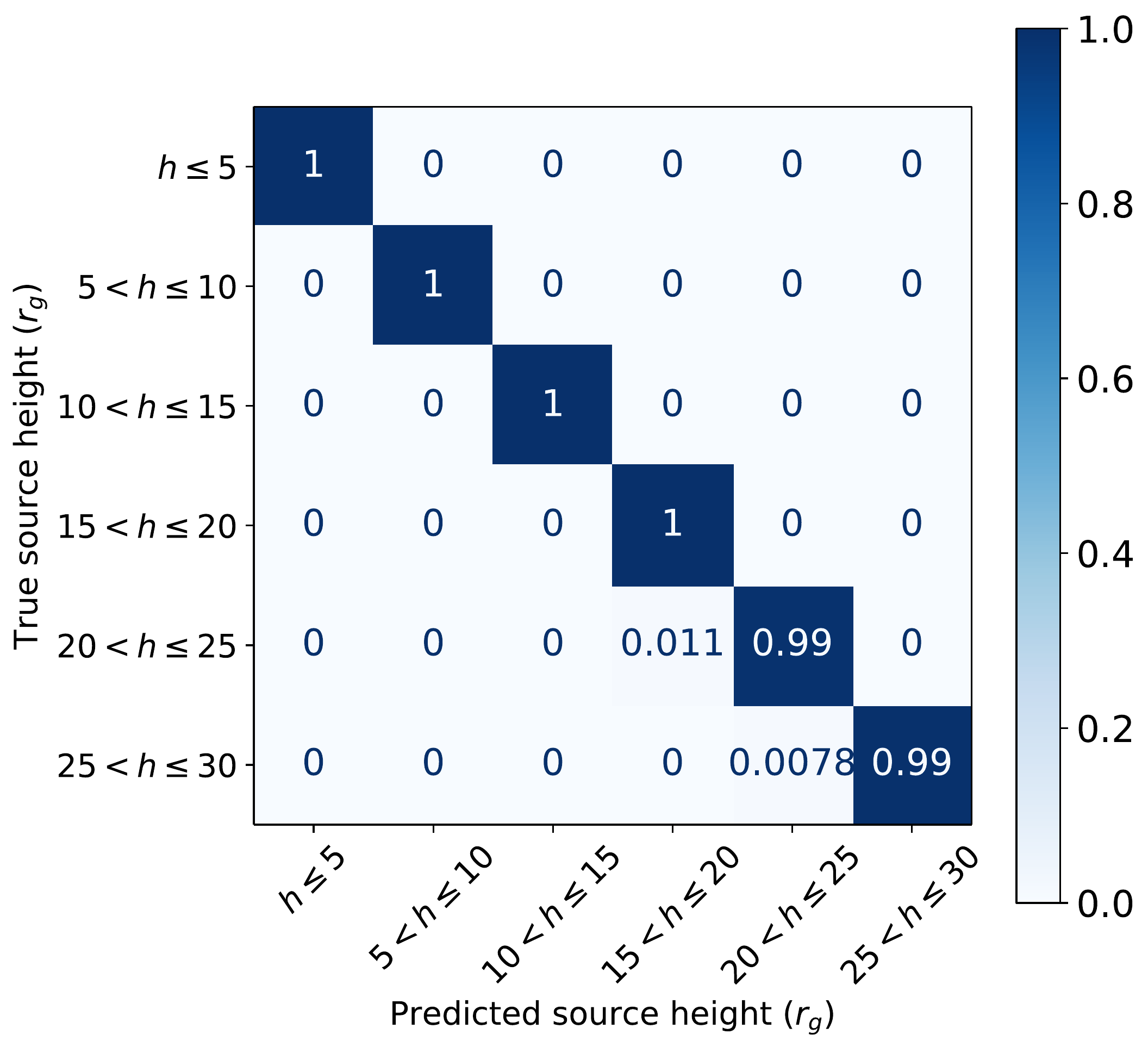}
    }
    \caption{Confusion matrix of the source height classification (predicted source height) compared to the true value (true source height) of the reconstructed PSD. The output numbers of samples are reported in each cell and are normalised over the sum of total samples scattered in the corresponding row. The mean accuracy of the classification is 0.99.}
    \label{fig-CM-h}
\end{figure}

We first classify the predicted results into 5 classes associated with 5 different ranges of source heights. The corresponding confusion matrix for the SVM classification of the source height using the reconstructed PSD in the test data set is shown in Fig.~\ref{fig-CM-h}. The mean accuracy of this classification is found to be $0.99$. Note that the accuracy reported here is the percentage of correctly classified samples out of all predictions. The cross-validation score and test score are high which also means that the model generalizes (extrapolates) well. 

\subsection{Classifying the source height and inclination}

\begin{figure}
    \centerline{
        \includegraphics[width=0.5\textwidth]{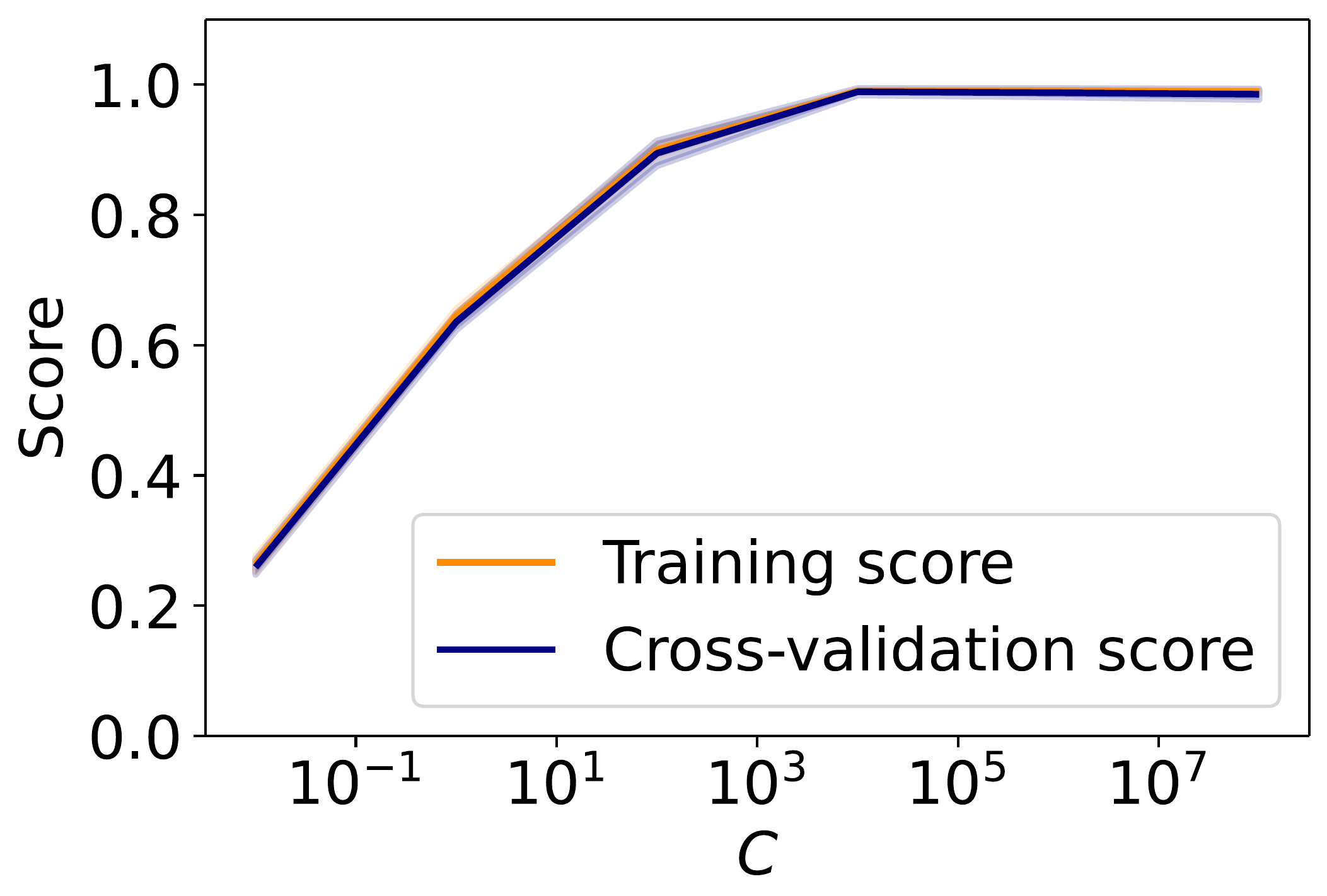}
    }
    \caption{Training scores (orange line) and validation scores (blue line) for classifying $h$ and $i$ for different hyperparameter $C$. High accuracy score is obtained when $C \gtrsim 10^4$.}
    \label{fig-C-h-i}
\end{figure}

\begin{figure}
    \centerline{
        \includegraphics[width=0.5\textwidth]{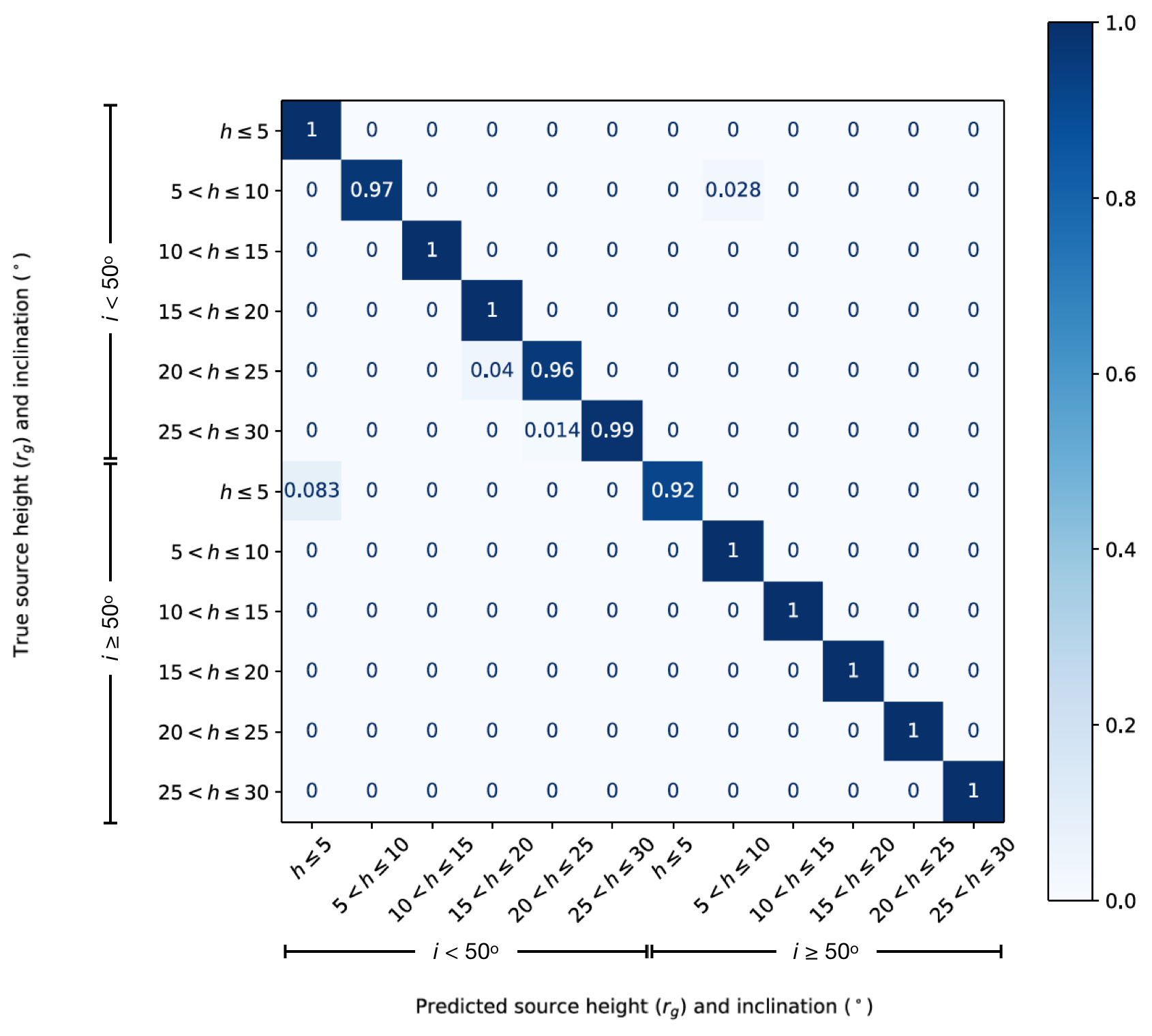}
    }
    \caption{Confusion matrix of the SVM classification for both source height and inclination. The mean accuracy of the classification is 0.98.}
    \label{fig-CM-h-i}
\end{figure}

Here, we evaluate the performance of our algorithm when both source height and inclination are placed as labels for the SVM classification. The machine is trained using the same parameter space as was previously used. Fig.~\ref{fig-C-h-i} shows the training and cross-validation scores for the SVM classification of $h$ and $i$ obtained during the hyperparameter tuning of linear kernel. $C=10^4$ is selected to be the best hyperparameter value. The corresponding confusion matrix is presented in Fig.~\ref{fig-CM-h-i}. The mean accuracy obtained in this case is 0.98. The prediction is slightly less accurate compared to when only the source height is classified alone.

\subsection{Generalization of ML}

\begin{figure}
    \centerline{
        \includegraphics[width=0.5\textwidth]{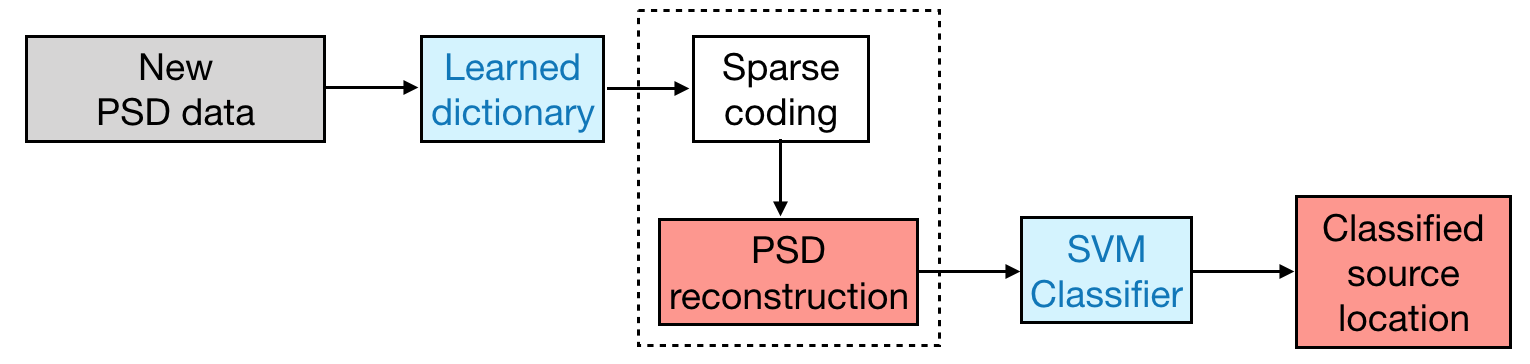}
    }
    \caption{Flowchart representing how the ML algorithm is applied for testing with new set of PSD data. This is to evaluate whether or not the trained machine is generalized well, so there is no additional training process implemented to the machine.}
    \label{fig-flowchart2}
\end{figure}

Now, we investigate further if the machine is capable of predicting the new data without needing to be re-trained. Therefore new data are directly applied to the ML algorithm as shown in
Fig.~\ref{fig-flowchart2}. The machine is said to be generalized well if the prediction on new data is accurate. 

\begin{figure}
    \centerline{
        \includegraphics[width=0.45\textwidth]{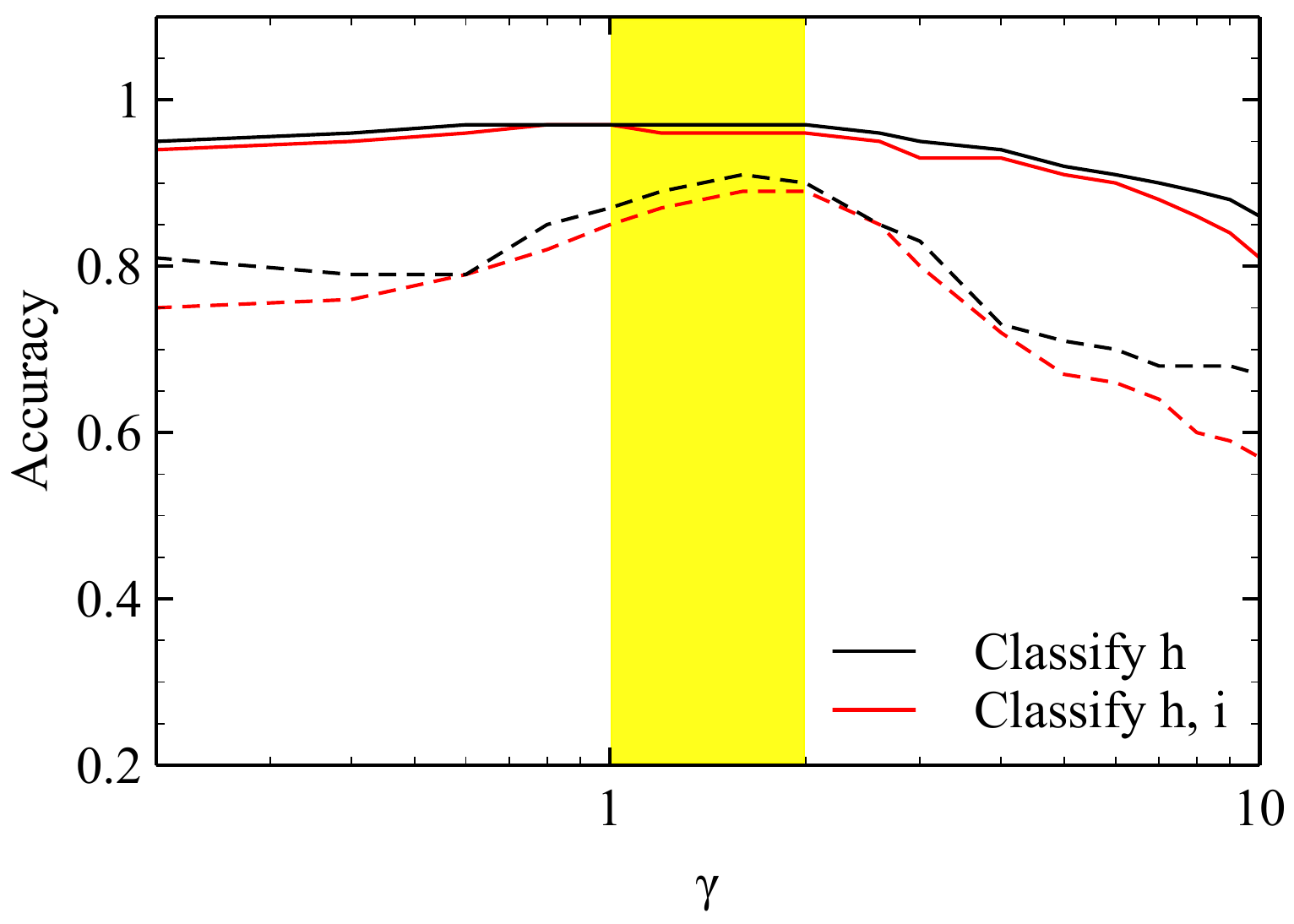}
    }
    \caption{Accuracy of the SVM classification of $h$ (black lines) and both $h$ and $i$ (red lines) depending on the power-law index $\gamma$ when $R_{\rm f}=0.7$. The dashed lines show the corresponding cases when $R_{\rm f}=0.1$. The yellow shade represents the range of $\gamma$ used to train the machine.}
    \label{fig-regularization}
\end{figure}

In Fig.~\ref{fig-regularization}, we demonstrate an accuracy of ML using the new PSD data having $\gamma$ between 0.2--10, with a given reflection fraction of 0.1 and 0.7 for comparison. Note that the machine is previously trained with $1.0 \leq \gamma \leq 2.0$. It can be seen that for $R_{\rm f}=0.7$, the ML accuracy is higher than 0.8 for all values of $\gamma$ even outside the range of which the machine is trained. The accuracy is high regardless of what option we are classifying: only $h$ or both $h$ and $i$. Interestingly, in the energy band containing significantly small amount of reflection flux of $R_{\rm f}=0.1$, the ML model can still make prediction with high accuracy of $> 0.75$ as long as $\gamma < 4.0$. Training the ML model using a wider range of parameter values will of course improve its accuracy, but we will not investigate further on this aspect which is not the main focus of this paper. Instead, we remark the benefit of the ML method that it can make a good accurate prediction of the source height even in the case when the contribution of the reflection flux is very low.  

Finally, we investigate the performance of the ML model by testing it with the new PSD profiles that are modelled in the form of a bending power-law:
\begin{equation}
P_{\rm 0}(f, E_{j}) \propto f^{-1}\bigg{[}1+ \bigg{(}\frac{f}{f_{b}} \bigg{)}^{s-1}\bigg{]}^{-1} \; ,
    \label{psd1}
\end{equation} 
where the PSD slope at low frequencies is fixed to $-1$ and $f_b$ is the bend-frequency above of which the PSD shape gradually bends to a slope $s$. We adopt the method of \cite{Timmer1995} to apply noises to the simulated bending power-law data. The noisy profiles are reconstructed via previously-trained dictionary before the corresponding source height is predicted using the SVM model. This is a data set that is completely new to the machine and is never used during the training phase. 

\begin{figure}
    \centerline{
        \includegraphics[width=0.45\textwidth]{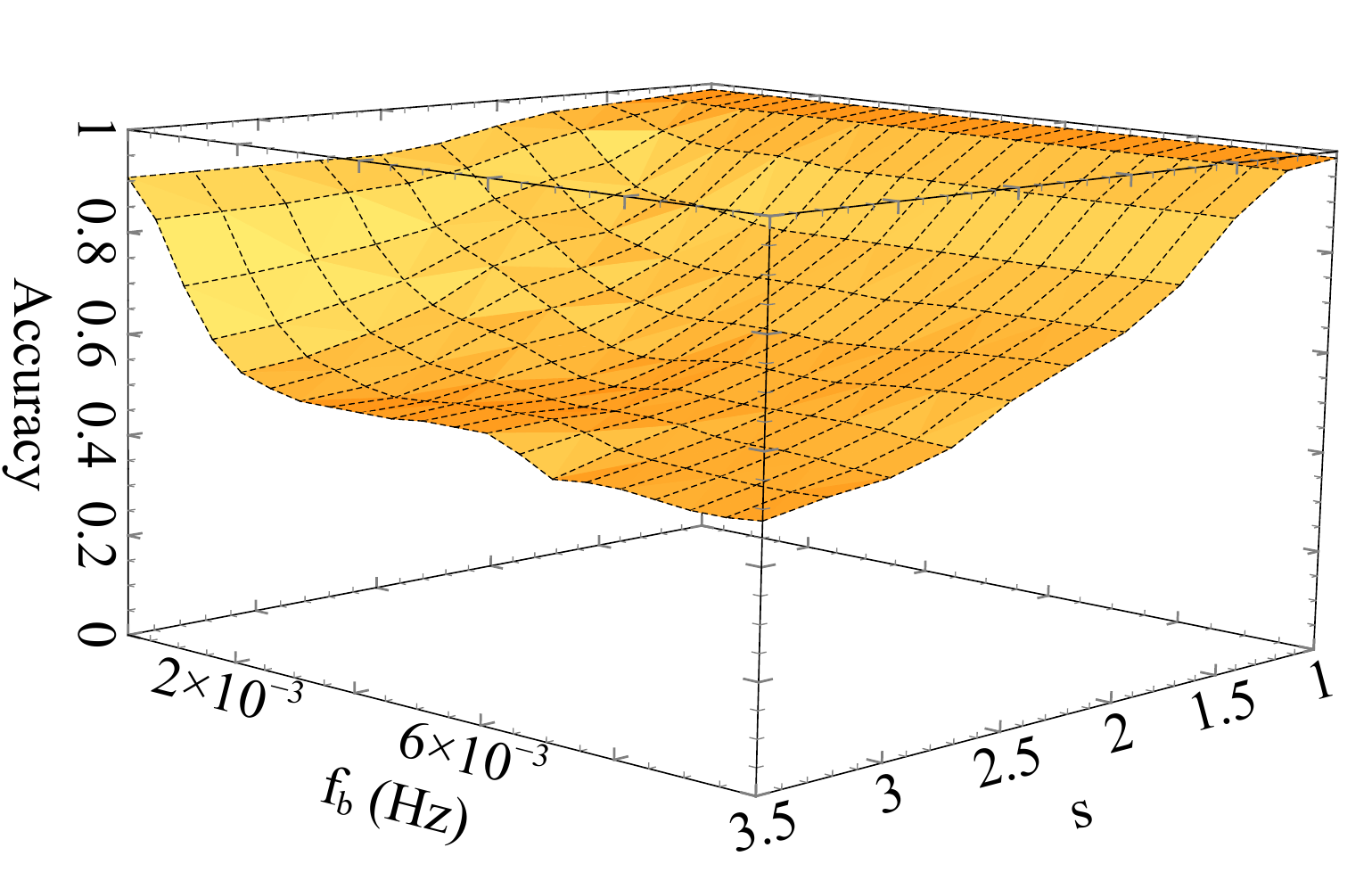}
    }
    \vspace{0.2cm}
    \centerline{
        \includegraphics[width=0.45\textwidth]{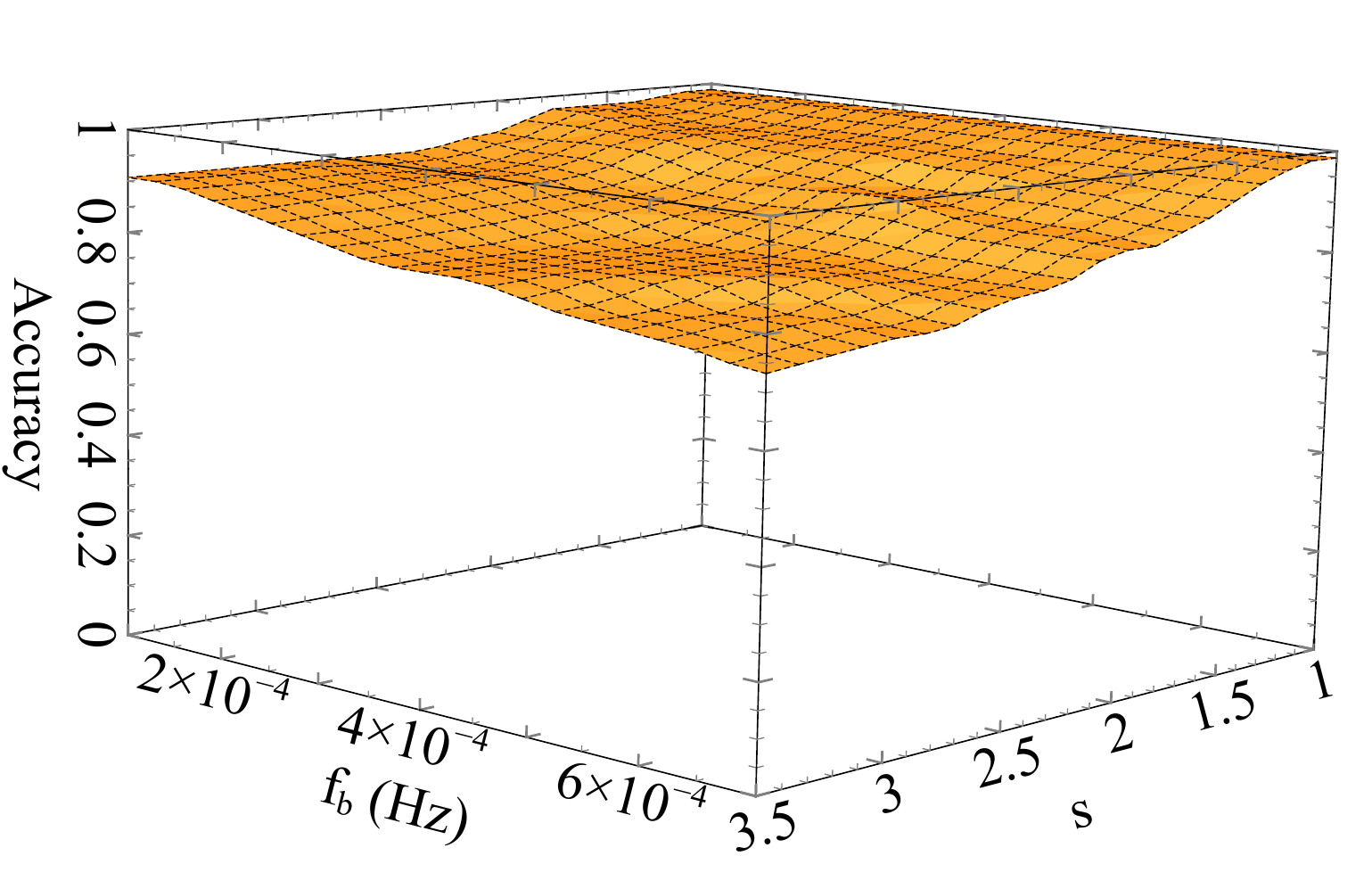}
    }
    \caption{Top pane: 2-D surfaces showing how the accuracy can be achieved on prediction the source height from new, bending power-law PSD data varying with the bend-frequency, $10^{-4} \leq f_b \leq 10^{-2}$~Hz, and the bend-index, $1.0 \leq s \leq 3.5$. We assume $R_{\rm f}=0.7$. Predictions are made by the machine trained with simple power-law model. It can be seen that the accuracy is high over a large range of parameters. Bottom panel: The same as top panel but focusing on $10^{-4} \leq f_b \leq 7 \times 10^{-4}$~Hz that covers the range usually observed in AGN.}
    \label{fig-test-bending-pwl}
\end{figure}

Dependence of the ML accuracy on the parameters $f_b$ and $s$ is presented in Fig.~\ref{fig-test-bending-pwl}. We fix $R_{\rm f}=0.7$ to represent the case when the reflection-dominated bands are used. The ML prediction gains significantly more accuracy when the bend-frequency is located at lower frequencies and when the bend-index is smaller. Fig.~\ref{fig-test-bending-pwl} (bottom panel) also shows the accuracy of source height prediction focusing on the parameter ranges of bending power-law model usually found in cases of AGN, which are  $1.0 \leq s \leq 3.5$ and $10^{-4} \leq f_b \leq 7 \times 10^{-4}$~Hz \citep[e.g.,][]{Emmanoulopoulos2016}. The accuracy score of $\gtrsim 0.75$ are obtained in these AGN parameter ranges even the profile shapes of these PSD data are deviated from what used to train the ML model.  

\section{Discussion}

In this work, ML algorithms are developed to identify the characteristic features of X-ray reverberation in PSD profiles containing information about the X-ray source location near black holes. Our scope of interest is to use only a fundamental PSD model to train the machine and evaluate how well the machine could learn and adapt to explain the unseen, more complex data.  

The observed PSD of AGN usually shows a great deal of scatter that normally has lower signal than the data in time-averaged spectra \citep[e.g.,][]{Martin2012}. It is therefore necessary to either tie some parameters in the reverberation model to those constrained from the time-averaged spectrum \citep[e.g.][]{Cackett2014, Wilkins2016, Epitropakis2016, Mahmoud2019, Chainakun2019b, Chainakun2021}, or perform simultaneous fits across different data sets \citep[e.g.][]{Chainakun2015, Chainakun2016, Mastroserio2020}. Here, we select to sparsely encode the noisy signals using the DL model that is trained with the clean PSD having a simple power-law shape. The simple but interpretable features on PSD are stored as the dictionary elements, called the atoms. During the reconstruction phase, the sparse coding is employed to activate a specific set of these atoms to reconstruct the nosiy PSD signals. Therefore transforming them to the sparse version whose relativistic echo features can be easily noticed and extracted by the SVM algorithm. In this way, the variability power on different temporal frequencies (i.e., different timescales) is considered as the features of the SVM model. We find that the most preferable ML model requires 64 features that are explained by 24 atoms.

According to Fig.~\ref{fig-CM-h}, the ML accuracy for locating the X-ray source could reach 0.99 if the source height is only the subject for ML prediction. Interestingly, the missing predictions occur only for the large source height of $h>20r_{\rm g}$, where the model seems to under-predict the true values. However, the number of these misclassified data are very small ($ \lesssim 2 \%$ of the total data). In fact, the error for misclassification of source height is within $\Delta h \sim 5r_{\rm g} $ which is, in some cases, smaller than the uncertainty of source height constrained using traditional timing analysis \citep[e.g.,][]{Emmanoulopoulos2014, Kara2016, Epitropakis2016, Emmanoulopoulos2016}. 

Furthermore, it can be inferred from Fig.~\ref{fig-CM-h} that the ML model succeeds in explaining the lamp-post heights regardless of the inclination angles. For determining the source height, it is not important whether or not the inclination is predetermined, e.g. by tying it to the value found in spectral fitting, in order to obtain the source location with high accuracy up to 0.99. We note that the test for cross-validation and training score have been also carried out (Fig.~\ref{fig-C-h} and Fig.~\ref{fig-C-h-i}). If the accuracy for classification of the test data is high but the cross-validation result is low, it means that the model performs well only in that particular group of data presenting in the test set (e.g., the model overfits to that data group only). Our test data in all cases are completely unseen in the way that they are never used during the training and validation process. The cross-validation results and test results, in turn, have high confidence all cases, suggesting that our ML models used in all analysis generalize (extrapolate) well. 

If we allow the ML to make predictions of both source height and inclination simultaneously, the obtained accuracy is also as high as 0.98 (Fig.~\ref{fig-CM-h-i}). Since we select an ensemble of only 5 inclinations, we label the data into two inclination classes, low and high, with the division at $50^{\circ}$, just for illustrating the ML performance. The machine can efficiently detect changes in reverberation feature due to inclination which are relatively small compared to changes by the effect of the source height. Also from Fig.~\ref{fig-CM-h-i}, there are some data associated with large source height whose values are under-predicted, as same as those found in Fig.~\ref{fig-CM-h}. However, it is clear that this under-estimation is not found in the case of high inclinations. Also, there is a very small number of misclassified data that lies far from the diagonal line of the confusion matrix. They are all height-label correct and their contributions constitute a minority. This is why the accuracy score is slightly lower than when the source height is predicted alone.  

Although the variations on the PSD slope across different energy bands are expected, the machine is trained to handle the PSD data with different power-law indices. The energy bands that have larger reflection fraction ($R_{\rm f}$) will have larger amplitude of the reverberation dip on the PSD profiles while the frequency of the dip remains the same \citep{Papadakis2016, Chainakun2019a}. Traditionally, the reflection fraction needs to be determined first, e.g. from spectral fitting, so that the lag and PSD data can be modelled with a proper dilution. We show in Fig.~\ref{fig-regularization} that the deviations in accuracy due to the choices of energy bands we chose to extract the PSD are less than $\sim 20\%$, compared between the cases of $R_{\rm f}=0.7$ and 0.1. Assuming the observed PSD data are close to the power-law shape with $\gamma < 3$, the ML accuracy in classification of its source height is higher than 0.7 even if the energy band that contains the reflection flux constitutes only 10\% of the total flux. If 70\% of the flux contributing in that energy band is due to reflection, the ML accuracy can be as high as $\sim 0.98$ when $\gamma < 3$. It also remains higher than 0.8 towards higher $\gamma$ beyond the upper limit of the values used in the training phase. 

Searching for X-ray reverberation signatures in the PSD of AGN was carried out by \cite{Emmanoulopoulos2016}. They found that strong constrains in the X-ray reflection geometry can be obtained using the current \emph{XMM-Newton} data if the intrinsic shape of the PSD is known in advance. We illustrate in Fig.~\ref{fig-test-bending-pwl} that the ML model developed here is capable of making predictions of AGN source height with high accuracy without needing to pre-determine the true shape of the observed PSD data. In fact, the accuracy is up to $\gtrsim 0.75$ if we scope down to the range of AGN parameters and becomes significantly greater for lower $f_{b}$ and $s$. This is because the bending PSD profiles become more similar to the simple power-law with the slope $-1$. Therefore, the shape of the test PSD is close to those used for training the machine. 

\begin{figure}
    \centerline{
        \includegraphics[width=0.45\textwidth]{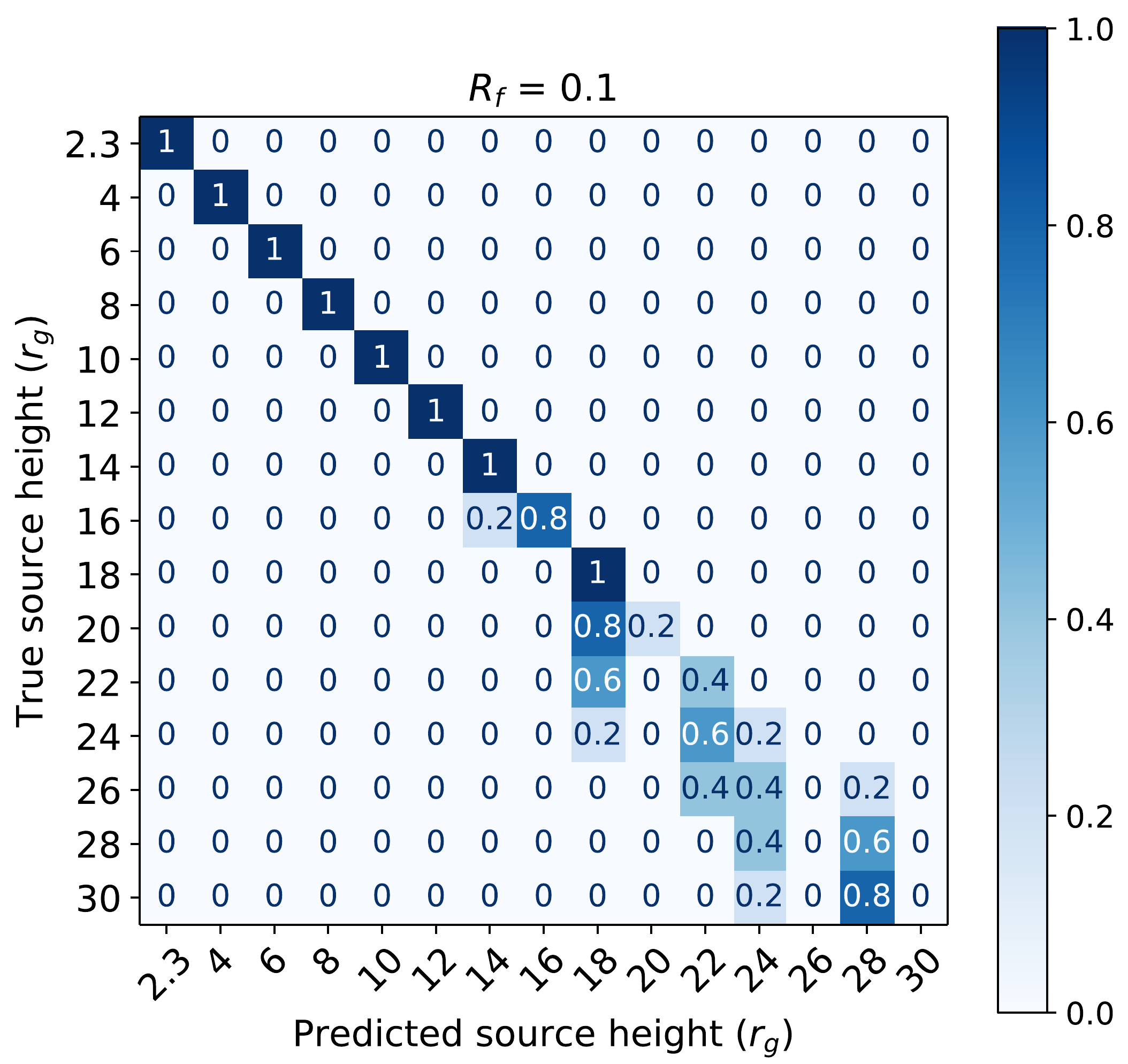}
    }
    \vspace{0.2cm}
    \centerline{
        \includegraphics[width=0.45\textwidth]{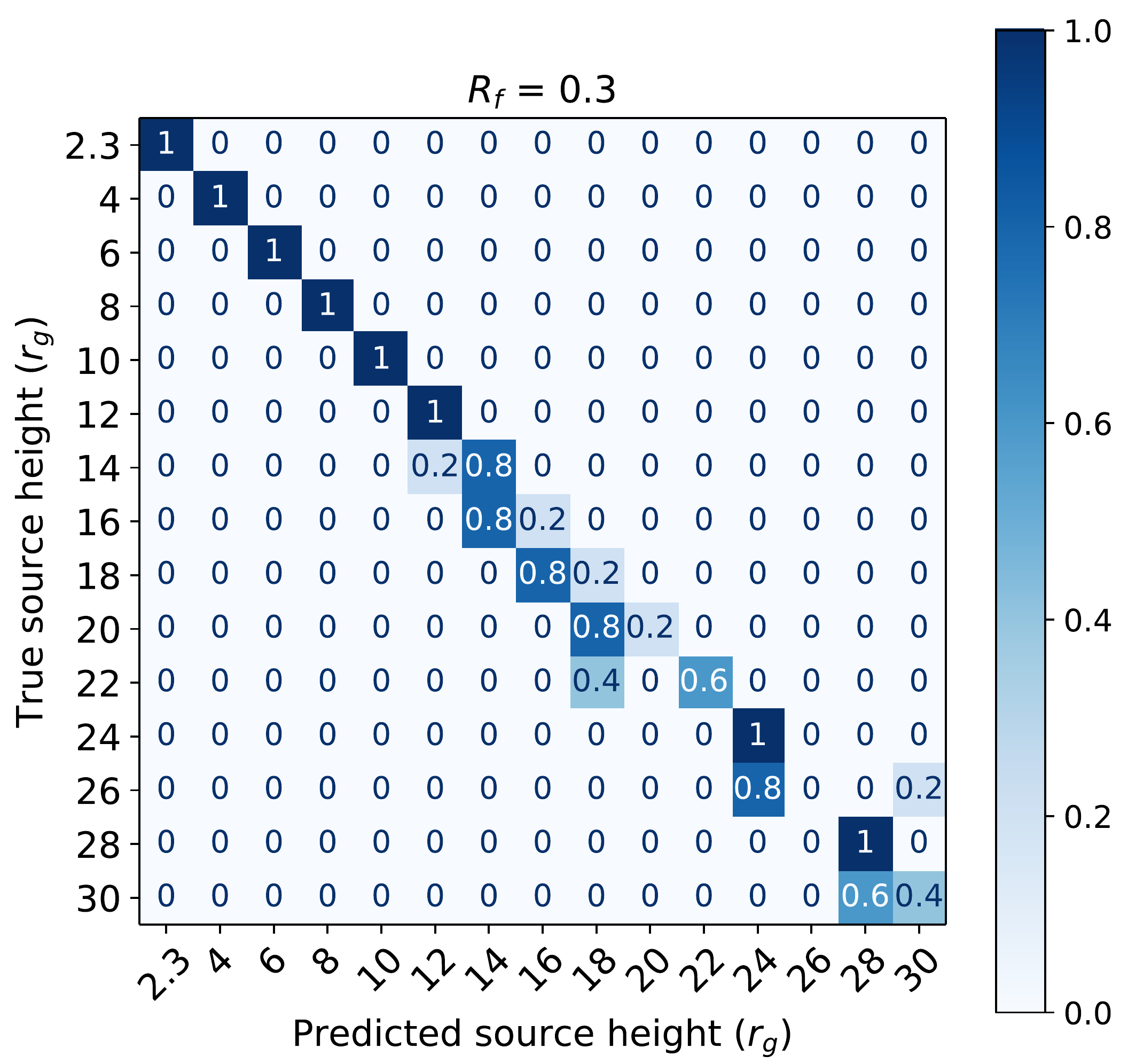}
    }
    \caption{Confusion matrix of the source height classification using all model grid of $h$ as labels when the reflection fraction $R_{\rm f}=0.1$ (top panel) and 0.3 (bottom panel). The test data are the bending power-law with the bend-frequency of $2 \times 10^{-4}$~Hz and the bend-index of $2$, the average values of which usually found in PSD of AGN. While these test data are completely new to the machine, the accuracy of predicting source height lower than $10 r_{\rm g}$ is 100\%. The deviation of predicted source height from the true values when $h > 10 r_{\rm g}$ is also small.}
    \label{fig-test-bending-pwl-2}
\end{figure}

\cite{Martin2012} studied the intrinsic shape of the PSD of 104 nearby AGN using the data is the \emph{XMM-Newton} archive. They reported that, for the majority of variable sources, their PSD could be explained by the bending power-law model with the mean index of $\sim 2$ and the mean bend-frequency of $\sim 2 \times 10^{-4}$~Hz. These mean values are of course small enough that the ML algorithm could make prediction of source height with very high accuracy. We elaborate this point by plotting the dependence of ML accuracy on predicting source heights for low reflection fractions of $R_{\rm f}=0.1$ and 0.3, when $f_{b}$ and $s$ are frozen at their mean values observed in AGN, as presented in Fig.~\ref{fig-test-bending-pwl-2}. It can be seen that, even in cases of such a low reflection fraction, the information of source height can be extracted with 100\% accuracy when the source height is below $10 r_{\rm g}$. The misclassification occurs only when the source height $h >10 r_{\rm g}$ and the deviation from the true values is small, with $\Delta h \sim 4 r_{\rm g}$ and $2 r_{\rm g}$ for $R_{\rm f}=0.1$ and 0.3, respectively. This small deviation of $\Delta h \sim 2 r_{\rm g}$ is also observed for larger $R_{\rm f}$ but with some (a certain) improvement on average accuracy. 

Furthermore, the deviations of the response functions may occur due to, e.g., the contribution of returning radiation \citep{Wilkins2020} or the effects of the accretion disc with a finite geometric thickness \citep{Taylor2018}. These phenomena could be manifested in the form of reprocessing echo structures in the PSD profiles. Although the classification results presented here are specific to the lamp-post geometry, the developed ML model should be capable of extracting reverberation features largely independent of the assumed disc-corona geometry they are produced from. Our results illustrate, as a proof of concept, that they can be beneficial for the X-ray reverberation studies in general. Wider ranges of ML models and assumptions are still open for further studies. For example, to identify reverberation features from the PSD of X-ray binaries where the intrinsic shapes may be more complicated with sometimes the presence of QPOs \citep[e.g.,][]{Veledina2016}. Furthermore, it is possible to fold the PSD in terms of the spectrograms, in term of studying the more complex features. In fact, there were previous studies in the literature suggesting that the DL technique would help studying the evolution of QPOs in X-ray binaries \citep{Lachowicz2010}. We do not convey this in our current work, but instead focus on a simple architecture of ML model which, in turn, can still provide a good prediction of the source height in AGN.  

Finally, the source height and the black hole mass could lead to the degeneracies of the model since they likely affect the reverberation lags in the same way \citep{Cackett2014, Alston2020}. AGN such as 1H~0707-495, Ark~564, MCG-6-30-15, and NGC~4051 that have black hole masses of the order of $\sim 10^{6} M_{\odot}$ \citep{Emmanoulopoulos2016} may be best suited to our current algorithm. Also, by fixing the black hole mass to known values reported in the literature, it can avoid the degeneracies in reverberation signals between the central mass and the source height, which are still not accounted for in our current investigation. Difficulties may arise due to the fact that any new data subject to be analyzed via the ML technique must be prepared in the same way as we did with the training data (e.g., the same normalizing and binning). Sufficiently detailed data may be required if one wants to train the regression model. We focus here only on the classification model that predicts classes assigned to different ranges of source heights, and the prediction has only right or wrong answers. However, the sense of how close the prediction is to the true class can be estimated from the confusion matrix. To handle a large number of observations and/or ranges of parameters, the most efficient procedure, regardless of whether one selects to treat the problem as classification or regression, may be to build a specific pipe-line that systematically transforms the observational data. Applying the ML technique to probe the X-ray source height using the AGN data in the \emph{XMM-Newton} archive will be reported in an upcoming paper.

\section{Concluding remarks}

In this work, we test the assumption that the X-ray reverberation features imprinted on the PSD data of AGN can be extracted using ML algorithms. We train the machine using only the fundamental power-law model, and expect it to make accurate predictions of AGN source heights for various types of PSD data. This assumption is specifically motivated by the foundation of ML that it can learn and adapt to new rules. We are not focusing on re-training the machine to get better accuracy, but on the performance of the developed ML to extract interpretable reverberation features. 

If the test PSD data are also in the form of a simple power-law, the prediction accuracy is up to $\sim 99\%$. The deviation of predicted source heights from the true values is found only in the case of $h>15 r_{\rm g}$, and only when the inclination is low (i.e., $i < 50^{\circ}$). The latter could be because larger $h$ produces smaller amplitude of the dip in the PSD profiles \citep{Papadakis2016} and lower inclination produces that the response functions get narrower in width due to smaller Doppler shift and relativistic beaming effects \citep{Cackett2014}. The interpretable PSD features for larger $h$ and lower $i$ then become more difficult to be distinguished by the ML model. The errors from misclassification in these cases, however, are very small. Even if we use an energy band that contains only 10\% of the reflection flux, the ML accuracy for predicting source heights is still high outside of the range of power-law indices for which the machine is trained. The accuracy can further increase if we use an energy band that is more dominated by the reflection flux. 

If the test PSD data exhibit a bending power-law shape, we find that the source height can still be located with the accuracy of $>0.75$ within the PSD parameter space of AGN reported in previous literature. Finally, we focus on the specific case of the bending power-law PSD having the bend-frequency and bend-index similar to the mean values usually seen in the real AGN data. We show that, in this case, only 10\% of reflection flux is required in any particular energy bands and it can be used by ML to extract the information of source height with 100\% accuracy for $h \leq 10 r_{\rm g}$. The misclassification is found only in cases of $h > 10 r_{\rm g}$, with small deviations from the true values of $\Delta h \sim 2$--$4r_{\rm g}$. 

Keeping in mind that the bending power-law data are completely new to the machine, this result suggests that the X-ray source can be located at high accuracy without needing to determine the intrinsic shape of the PSD to train the machine in advance. Although our results may be specific to the scope of investigation and assumptions made here, it suggests that the ML technique could contribute to new directions for methodological development in time series analysis. In particular, it could be an alternative powerful way to trace the reverberation signatures. Developing a pipe-line supporting the test of the ML model on real PSD data in the \emph{XMM-Newton} archives is planned for the future.

\section*{Acknowledgements}
The ML models in this work are adopted from the {\sc sklearn} software package. N.M. and P.T. thank for the financial support from Suranaree University of Technology. We thank the referee for helpful comments which have improved the paper.

\section*{Data availability}
The response function data analysed here are generated using the {\sc kynxilrev} model available in \url{https://projects.asu.cas.cz/stronggravity/kynreverb}. To analyse the PSD data, we adopt the ML models available in {\sc scikit-learn} at \url{https://scikit-learn.org/}. The derived data underlying this article will be shared on reasonable request to the corresponding author.


\bibliographystyle{mnras}

\bsp	
\label{lastpage}
\end{document}